\documentclass[twocolumn, twocolappendix]{aastex631}
\usepackage{color}
\usepackage{natbib}
\usepackage{amsmath}
\usepackage{appendix}
\usepackage[utf8]{inputenc}
\usepackage{longtable}
\usepackage{threeparttablex}
\usepackage{multirow}
\usepackage{footmisc}

\newcommand{\T}{\mathbf{T}}
\newcommand{\degrees}{^\circ}
\newcommand{\angstrom}{\text{\normalfont\AA}}
\newcommand{\msun}{M$_{\odot}$}

\newcommand{\kms}{\rm{km}~\rm{s}^{-1}}
\newcommand{\wlocalv}{$W_{v, \rm{local}}$}
\newcommand{\wlocalx}{$W_{\theta, \rm{local}}$}
\newcommand{\wglobalv}{$W_{v, \rm{global}}$}
\newcommand{\wglobalx}{$W_{\theta, \rm{global}}$}

\newcommand{\HI}{H{\sc\ i}}

\newcommand{\SiII}{Si{\sc\ ii}}
\newcommand{\MgII}{Mg{\sc\ ii}}
\newcommand{\SiIII}{Si{\sc\ iii}}
\newcommand{\SiIV}{Si{\sc\ iv}}
\newcommand{\SII}{S{\sc\ ii}}
\newcommand{\CII}{C{\sc\ ii}}

\newcommand{\CIV}{C{\sc\ iv}}
\newcommand{\OVI}{O{\sc\ vi}}

\shortauthors{Kim et al.}

\begin{document}
\defcitealias{Karachentsev13}{K13}
\defcitealias{putman21}{P21}
\defcitealias{richter17}{R17}
\defcitealias{fox20}{F20}
\defcitealias{Krishnarao22}{K22}
\defcitealias{lehner20_amiga}{L20}
\defcitealias{Zheng_ic1613}{Z20}
\defcitealias{Nidever08}{N08}

\title{Identifying H{\sc\ i} Emission and UV Absorber Associations Near the Magellanic Stream}

\author[0000-0001-9654-5889]{Doyeon A. Kim}\thanks{E-mail: d.kim3@columbia.edu}
\affiliation{Department of Astronomy, Columbia University, New York, NY 10027, USA}

\author[0000-0003-4158-5116]{Yong Zheng}
\affiliation{Department of Physics, Applied Physics and Astronomy, Rensselaer Polytechnic Institute, Troy, NY 12180}

\author[0000-0002-1129-1873]{Mary E. Putman}
\affiliation{Department of Astronomy, Columbia University, New York, NY 10027, USA}

\date{\today}

\begin{abstract}
We present a new technique to identify associations of \HI\ emission in the Magellanic Stream (MS) and ultraviolet (UV) absorbers from 92 QSO sight lines near the MS. We quantify the level of associations of individual \HI\ elements to the main \HI\ body of the Stream using Wasserstein distance-based models, and derive characteristic spatial and kinematic distances of the \HI\ emission in the MS. With the emission-based model, we further develop a comparison metric, which identifies the dominant associations of individual UV absorbers with respective to the MS and nearby galaxies.
For ionized gas associated with the MS probed by \CII, \CIV, \SiII, \SiIII, \SiIV, we find that the ion column densities are generally $\sim$0.5 dex higher than those that are not associated, and that the gas is more ionized toward the tail of the MS as indicated by the spatial trend of the \CII/\CIV\ ratios. For nearby galaxies, we identify potential new absorbers associated with the CGM of M33 and NGC300, and affirm the associations of absorbers with IC1613 and WLM. For M31, we find the previously identified gradient in column densities as a function of impact parameter, and that absorbers with higher column densities beyond M31's virial radius are more likely to be associated with the MS. 
Our analysis of absorbers associated with the Magellanic Clouds reveals the presence of continuous and blended diffuse ionized gas between the Stream and the Clouds. Our technique can be applied to future applications of identifying associations within physically complex gaseous structures.
\end{abstract}

\keywords{Magellanic Clouds (990); Magellanic Stream (991); Milky Way Galaxy (1054); Circumgalactic medium (1879)}

\section{Introduction}
A diffuse gaseous medium comprises the majority of the baryonic mass in the Universe and spans a wide range of dynamic and energy scales \citep{hi_dickey, putman_halogas, tumlinson_cgm, multiphase_peroux}. A comprehensive identification and categorization of gaseous media is challenging, as it is found in versatile forms \citep{heiles_hi_shape}, and is involved in various physical processes \citep{jerry_mckee_ism, Hartquist94_ism}. Nonetheless, the direct association of a gaseous medium with a nearby galactic structure or environment is also difficult, because observations can only provide a limited projection of its true nature.

Both absorption and emission lines offer effective ways to probe the diffuse gas in galaxy halos. The absorption lines found in the spectra of bright background objects, such as quasars (QSOs), efficiently probe low-density gas at various ionization states with a broad range of metal transition lines \citep{gunn_peterson, bahcall_spitzer, schaye_review, tumlinson_cgm}. However, the limited spatial coverage of a QSO pencil-beam sight line makes it challenging to associate an absorber system with its ambient environment. Typically, an absorber is considered to be associated with a galaxy when it is in ``close" proximity to the galaxy in position and velocity space. Considering an impact parameter ($b$) as a proxy for a galaxy's influence on its surrounding gas environment, the circumgalactic medium (CGM) has been probed with QSO sight lines located within certain impact parameter thresholds \citep[e.g.][]{chen_tinker_abs, prochaska_abs, Werk16_hot_halo, rudie19}. 

The appropriate choice for an impact parameter threshold to distinguish a galaxy's CGM from the intergalactic medium remains somewhat elusive. \cite{tumlinson13_hi}, for instance, demonstrated the presence of a substantial gas reservoir with a nearly 100$\%$ \HI\ covering fraction for star-forming galaxies and $\sim$75$\%$ for passive galaxies out to 150 kpc. Other studies found that extended \HI\ halos can be detected out to impact parameters of $\sim$300 kpc in various types of galaxies at low redshifts (z$<$1)
\citep{chen98_absorption, bowen02, keeney17}. Moreover, \cite{liang_rvir} detected halo gas reservoirs extending out to 500 kpc with a mean covering fraction of 60\%, and it has been found that the gaseous medium around a galaxy can extend up to $\sim1$ Mpc for weak absorbers \citep{morris93_absorption, tripp98_absorption, Wakker09_absorption, Prochaska11, tejo14_absorption}. \cite{cgm_hi_wilde} found that the appropriate impact parameter threshold to associate an absorber with a galaxy depends on the galaxy's mass. Furthermore, low ions, such as \MgII\ and \SiII, are rarely detected at impact parameters greater than 100 kpc from their host galaxies \citep{bordoloi11_mg2, farina13, werk13_mg2}, while highly ionized metal absorbers such as \OVI\ are detected out to $\sim$100--500 kpc from the host galaxy \citep{tumlinson11_ovi, bielby_ovi, lehner20_amiga, Tchernyshyov23}.

\begin{figure*}[t]
    \centering
    \includegraphics[width=\linewidth]{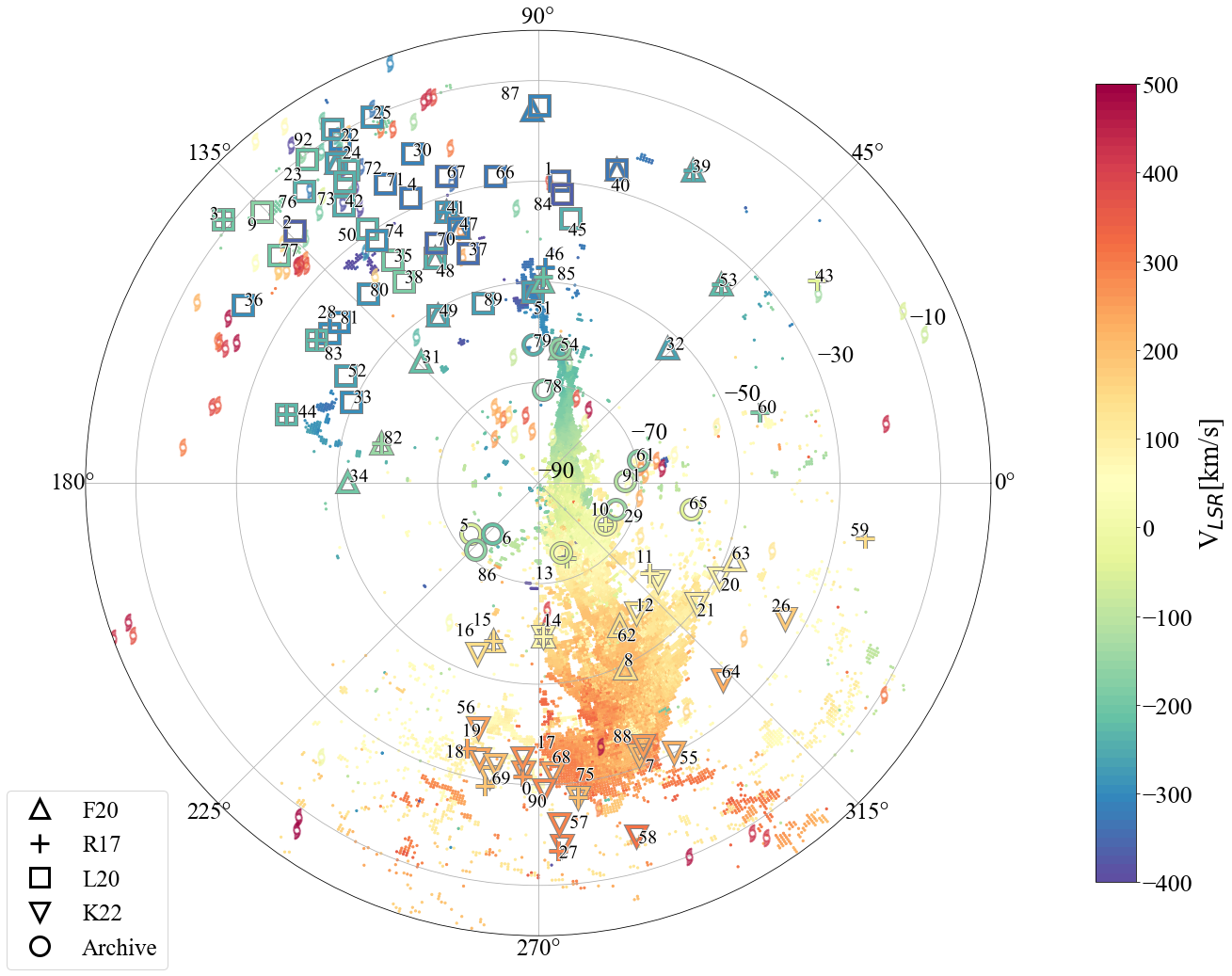}
    \caption{The MS \HI\ data \citep[background dotted patches;][]{Nidever08}, our QSO sight lines (symbols shown in the lower left legend; Section \ref{sec:qso_data}), and nearby galaxies with M$_{\rm HI} > 10^5$ \msun\ \citep[galaxy markers;][]{Karachentsev13, putman21} at Galactic latitude of GLAT $\leq 0\degrees$. All data    are plotted in Galactic coordinates, and color-coded based on their radial velocities in the LSR frame. We only include galaxies with radial velocities within $\pm 500 \kms$ and saturate the color scale to cover [-400, 500] $\kms$ for illustration purposes.
    For QSO sight lines where multiple absorbers are present, the colors represent the averages of all absorber velocities along the corresponding sight lines. 
    The number next to each symbol indicates the ID assigned to the corresponding QSO sight line in Table \ref{tb:qso}. The shapes of the symbols indicate their data origins, with up-pointing triangles for \citet[][F20]{fox20}, down-pointing triangles for \citet[][K22]{Krishnarao22}, crosses for \citet[][R17]{richter17}, squares for \citet[][L20]{lehner20_amiga}, and circles for our archive sample (see Section \ref{sec:data} for more details).}
    \label{fig:MS_full}
\end{figure*}

Emission line data offer a comprehensive picture of the spatial and kinematic extent of a gaseous medium. Emission observations, like 21 cm \HI\ measurements, are typically reported in position–position–velocity (PPV) space, where the third dimension represents the line-of-sight velocity. A wider spatial view and additional kinematic information often makes an association more straightforward. Nevertheless, emission structures, particularly within the interstellar medium (ISM), can overlap, making it challenging to disentangle associated emissions. Although recent work infers three-dimensional distribution and association of the Milky Way's ISM structures \citep{reach15, capitanio, zucker18_co, panopoulou20} with high accuracy astrometric and photometric surveys of stars \citep{panstarr, gaia}, such association is dependent on stellar  locations. Establishing the broader spatial and kinematic association of emission line data still remains critical for a relevant ISM research.

Associating emission with various structures has been performed by different groups. Typically, the emission line structures are associated based on similarities in kinematics or thermal properties and their proximity in position and velocity space \citep{cox74, MC02, haud_gc, Nidever08,  goodman09, gaskap_mc, kim23}. Some approaches employ physical models of expected emission from a structure to associate the observed emission \citep{pikelner68, blitz79,wakker91,westmeier18}. With each group optimizing their approach for specific applications, the current-state-of-art lacks standardization. A way to standardize the association is through quantification of related metrics. For example, \cite{Peek_Dparam} quantified the level of association between a cloudlet and a nearby high-velocity cloud complex by defining a distance metric in position-velocity space and modeling the association based on simulated clouds. Since this association metric is based on a specific simulation, it may not be applicable to identify gaseous associations more broadly. The challenge of quantifying the likely association among gaseous structures escalates for a complex system like the Magellanic Stream. 

The Magellanic Stream (MS) is a massive \HI\ structure located in the halo of the Milky Way (MW) and connected to the Large and Small Magellanic Clouds (LMC and SMC). The MS is composed of multiphase gas with distinct kinematic and chemical properties \citep{Sembach2003, mcclure_hvc_mwdisk, richter17, fox20}, including two bifurcated \HI\ filamentary structures that are possibly linked to the LMC and SMC \citep{putman2003, Nidever08}. Despite extensive observations and simulations, the exact origin and physical mechanisms shaping the MS remain unclear. 
Tidal forces between the LMC and SMC, as well as their interaction with the MW, have been proposed to explain the existence of the MS and its extended multiphase gaseous features \citep{lin95, Connors06, besla12, diaz12}. Ram pressure forces exerted on the LMC and SMC as the galaxies travel through the MW's CGM may also contribute to the origin and present-day appearance of the MS \citep{salem15,tepper15}. 
Recent observational evidence on the existence of a Magellanic corona \citep{lucchini20,Krishnarao22}, a Magellanic stellar stream \citep{Chandra23}, and additional energetic processes near the Galactic Center \citep{BH13,barger_halpha, fox20} further complicates the interpretation of the gaseous multiphase structure in the vicinity of the MS.

In this work, we use both ultraviolet (UV) absorption and neutral hydrogen (\HI) 21cm emission line observations to quantify the associations of gaseous medium near the MS and nearby galaxies in close projection. In Section \ref{sec:data}, we review the data that are used in this work. In Section \ref{sec:distance_model}, we first build an association metric based on the \HI\ emission of the MS using a Wasserstein distance. Based on the association metric, we quantify the levels of associations of individual \HI\ elements in the MS. We further develop this association metric to examine the associations of UV absorbers with the MS or nearby galaxies in Section \ref{sec:assoc_eval}. In Section \ref{sec:absorber_result}, we present absorber associations, compare our associations with previous studies in the literature, and discuss the physical conditions of the MS. In Section \ref{sec:discussion}, we compare our association metric to other existing metrics and further discuss the physical implications of our work on the MS and nearby galaxies.

\section{Data} \label{sec:data}
\subsection{\HI\ Emission Data}
\label{sec:hi_data}
We obtain the \HI\ emission data from \citet[][hereafter, \citetalias{Nidever08}]{Nidever08}, which is primarily based on the Leiden/Argentine/Bonn (LAB) all sky \HI\ survey \citep{LAB}. The LAB all-sky \HI\ survey is a combined dataset between the Leiden/Dwingeloo Survey that covers the northern sky over declination ${\rm Decl.}> -30\degrees$ \citep{LDS} and the Instituto Argentino de Radioastronomı´a Survey that covers ${\rm Decl.}\leq -25\degrees$ \citep{IAR2, IAR1}. The combined dataset has a velocity and spatial resolution of $1.3~\rm{km}\rm{s}^{-1}$ and 36$'$, respectively. The velocity range is from $-400$ to 400 $\rm{km}\rm{s}^{-1}$ and its RMS noise is $0.09$ K. 

With the LAB survey, \cite{Nidever08} decomposed the \HI\ emission in the vicinity of the MS into Gaussian components and grouped the \HI\ emission based on the phase structures of the gas \citep{field69, jerry_mckee_ism, Ferriere01}. Their Gaussian decomposition disentangles the MS from the MW gas and preserves the continuity of MS's \HI\ filamentary structures in position and velocity space. For each \HI\ data point from \citetalias{Nidever08}, which we define as an \HI\ element, the Gaussian decomposition provides the amplitude of the \HI\ emission, the centroid velocity $v_{\rm HI}$ in the local standard of rest (LSR), and the velocity dispersion $\sigma_{\rm v, HI}$. Our distance model will be based on the ($v_{\rm HI}$, $\sigma_{\rm v, HI}$) measurements from \cite{Nidever08}, which we elaborate on in Section \ref{sec:distance_model}.

\subsection{Nearby Galaxies}
\label{sec:galaxy_data}
We utilize the all-sky catalogs of nearby galaxies within the Local Volume provided by \citet[][hereafter \citetalias{Karachentsev13}]{Karachentsev13} and \citet[][hereafter \citetalias{putman21}]{putman21}. We work with an underlying assumption that if a galaxy in the Local Volume does not have detectable \HI\ gas in its main body, then it is unlikely to have an associated gaseous absorption line in its CGM.  We adopt the galaxies detected in \HI\ by \citetalias{putman21} and additional galaxies from the \citetalias{Karachentsev13} catalog whose \HI\ gas mass is above $10^{5}$ \msun\ and within the velocity range $\pm$ 500$\kms$. The \citetalias{Karachentsev13} catalog encompasses various observed properties of nearby galaxies, including apparent magnitudes, K-band luminosities, \HI\ fluxes, radial velocities, distances, and derived properties such as \HI\ masses and M26, which are the dynamical masses within Holmberg radii. In cases where there are discrepancies in the reported \HI\ masses between the \citetalias{Karachentsev13}'s and \citetalias{putman21}'s catalogs, we preferentially rely on the values from \citetalias{putman21}. 
Figure \ref{fig:MS_full} displays all galaxies that surpass the \HI\ mass threshold and are located at Galactic latitudes less than zero degrees (GLAT$ \leq 0 \degrees$).

\subsection{UV Absorbers from QSO Sight Lines}
\label{sec:qso_data}
We compile our UV absorber dataset from five sources. We begin with recently observed data from our Hubble Space Telescope (HST) program (PID:16301) and conduct a search in the Mikulski Archive for Space Telescopes (as of November 2022) for QSO sight lines within 1000 kiloparsecs (kpc) of the Sculptor group; hereafter, we refer to this set of data as the archive sample. Further details on our archive sample regarding the sight line selection, spectral co-addition, and absorption line analyses can be found in Appendix \ref{appendix:archive}. 

We also compile sight lines from four published works, including (1) \citet[][hereafter, \citetalias{richter17}]{richter17} that characterized the UV absorption of high-velocity clouds in the MW and Local Group toward 270 sight lines, (2) \citet[][hereafter, \citetalias{fox20}]{fox20} that modeled the kinematic structure of ionized absorbers near the MS (within 30$\degrees$) toward 31 sight lines, (3) \citet[][hereafter, \citetalias{Krishnarao22}]{Krishnarao22} that examined the ambient medium of the LMC within 35 kpc for evidence of a Magellanic corona using a sample of 28 sight lines located within 45$\degrees$ from the LMC, and (4) \citet[][hereafter, \citetalias{lehner20_amiga}]{lehner20_amiga} that surveyed 43 sight lines at impact parameters ($b$) between 25 to 569 kpc from M31. \citetalias{lehner20_amiga} further identified the MS absorbers to be located within 20$\degrees$ of the MS latitude. From these literature works, we only select sight lines with UV absorbers that are near the MS and located at GLAT$ \leq 0 \degrees$. 

The unique aspect of this work is that we implement a comprehensive statistical technique to evaluate the UV absorber associations with the MS using both kinematic and spatial information, and we identify the associations \textsl{quantitatively}. This is in contrary to the aforementioned UV studies, which mainly used simple cuts in sky locations and velocities to identify absorber associations.

All QSO sight lines that we have compiled, including our archive sample and literature values, were observed with the Cosmic Origins Spectrograph (COS) on board the HST with G130M and/or G160M gratings, covering a wavelength range of $\lambda \sim $1150--1450$\angstrom$ and $\lambda \sim $1405--1775$\angstrom$, respectively. For UV absorber measurements used in this work, we denote the absorbers' centroid velocities in the local standard of rest (LSR) as $v_{\rm ion}$. We only collect absorbers with $v_{\rm ion}$ within a velocity range of [-500, 500] $\kms$.

We focus on a set of common absorption lines from silicon and carbon ions, including \SiII, \SiIII, \SiIV, \CII, and \CIV. More specifically, we consider the following UV transition lines: \CII\ $\lambda$1334, \SiII\ $\lambda$1193, \SiIII\ $\lambda$1206, \SiIV\ $\lambda$1393, and \CIV\ $\lambda\lambda$1548. While many of these ions have multiple transition lines (e.g., the 1393/1402 Å doublet for \SiIV), we mostly adopt the stronger lines, which yield higher signal-to-noise ratios. For \SiII, we choose a weaker \SiII\ $\lambda$1193 line instead of $\lambda$1260, because \SiII\ $\lambda$1260 is often heavily saturated and blended with \SII\ $\lambda$1259 from potential high-velocity clouds in the MW along the same line of sight. Following \cite{fox14}, $\lambda$1260 is used in lieu of \SiII\ $\lambda$1193 at positive absorption velocities where the MW's \SII\ $\lambda$1259 is not a concern. When certain lines are either saturated or contaminated, we instead take the other available transition lines of the same ions, such as \SiII\ $\lambda$1526, \SiIV\ $\lambda$1402, or \CIV\ $\lambda\lambda$1550.

Table \ref{tb:qso} presents the absorber data for a total of 870 UV absorbers collected from 92 QSO sight lines near the MS. Amongst the 870 absorbers, 372 are from \citetalias{lehner20_amiga}, 242 from \citetalias{Krishnarao22}, 106 from \citetalias{fox20}, 76 from the archive, and 74 from \citetalias{richter17}. We assign a numerical ID to each sight line and show the sight lines with the corresponding IDs
in Figure \ref{fig:MS_full}.
For the QSO sight lines in our archive sample, we calculate the column density of each absorber using the apparent optical depth (AOD) method \citep{Savage91}, consistent with the method adopted in \citetalias{richter17} and \citetalias{lehner20_amiga}. Meanwhile, \citetalias{fox20} and \citetalias{Krishnarao22} measured their absorber column densities by identifying distinct velocity components in the absorption line data and fitting them with Voigt profiles (VP). 
The different techniques may result in variation in the assessment of an absorber's column density and velocity range. We elaborate on the main differences between the AOD and VP methods as follows, and discuss our approach to combine the datasets from different sources. 

The AOD-based method as adopted by \citetalias{richter17}, \citetalias{lehner20_amiga}, and this work (the archive sample) identifies absorption velocity ranges mainly through visual inspection, and calculates total column densities by integrating the normalized line profiles over the designated velocity ranges using equation \ref{eq:aod_column}. The \citetalias{richter17} dataset and our archive sample identify all metal absorbers that occur over a velocity range from $v_{\rm LSR}\sim-500~\kms$ to $\sim+500~\kms$. \citetalias{lehner20_amiga} instead focused on $ -510 \lesssim v_{\rm LSR} \lesssim -150~\kms$ for absorption related to M31; we adopt \citetalias{lehner20_amiga}'s full dataset without considering the M31 or MS membership that they assign to their absorbers. In Section \ref{sec:absorber_result_compare}, we compare our association results to \citetalias{lehner20_amiga}'s association result.

For the VP-based method, \citetalias{fox20} modeled both the low ($\rvert v_{\rm{LSR}} \lvert < 100~\kms$) and high ($\rvert v_{\rm{LSR}} \lvert  > 100~\kms$) velocity components simultaneously in their VP fits to better constrain the physical parameters of the MS-associated components. 
In contrast, \citetalias{Krishnarao22} exclusively considered absorbers with $\rvert v_{\rm{LSR}} \rvert > 150~\kms$ to exclude absorbers with velocities associated with known MW intermediate-velocity and high-velocity clouds in the region \citep{wakker_hvc, putman_halogas}.

In Table \ref{tb:qso}, for absorber data calculated based on the AOD method, we adopt the minimum and maximum velocity ranges ($v_{\rm min}$, $v_{\rm max}$) of the absorbers directly from the corresponding references. For those that are based on the VP method, we first convert each velocity component's Doppler width to the 1D velocity dispersion ($\sigma_{\rm v, ion}$) as $\sigma_{\rm v, ion} \equiv {\rm Dopper~width}/\sqrt{2}$ \citep{Draine_ism}; for these absorbers, the ($v_{\rm min}$, $v_{\rm max}$) values in Table \ref{tb:qso} reflect the extents of the velocity components' full-width-half-maximum (FWHM $\equiv 2.355\sigma_{\rm v, ion}$). Furthermore, all signal-to-noise ratios (S/N) per resolution element listed in Table \ref{tb:qso} are directly sourced from the original literature. \citetalias{richter17} calculated their S/N averaged over an absorption-free spectral range of 1208--1338 \AA, \citetalias{fox20} estimated theirs near \SII\ 1250 \AA, and \citetalias{lehner20_amiga} estimated theirs near \SiIII\ 1206 \AA. The S/N values of the sight lines in our archive sample are estimated over a wide spectral range of 1150--1750 \AA\ (see \citealt{zheng23}). As the \citetalias{Krishnarao22} sample did not provide S/N values for their published absorbers, 
we leave these entries empty in Table \ref{tb:qso}. We also exclude absorbers from \citetalias{Krishnarao22} that yield what they refer to as poor component fits. We adopt absorbers from the four published literature references as they are and do not implement S/N cuts in these data.

We cross-match the QSO sight lines among the five sources (our archive sample, \citetalias{richter17}, \citetalias{lehner20_amiga}, \citetalias{fox20}, and \citetalias{Krishnarao22}) and identify the common sight lines that are included in more than one study. For these sight lines, measurements of their absorption line features may vary among different works depending on the particular methods being used. For absorbers of the same ions with centroid velocities within the COS resolution limit ($\approx 25~\kms$) from different works \citep{cos_handbook}, we favor the VP-based data over the AOD-based data. If an absorber appears in both \citetalias{fox20} and \citetalias{Krishnarao22}, which both applied VP fitting in their analyses, we prioritize the measurement from \citetalias{fox20} for its known S/N. Similarly, if multiple AOD data exist for the same absorber, we adopt the most recent measurement in the order of our archive, \citetalias{lehner20_amiga}, and then \citetalias{richter17}.

By design, the absorbers that we compile in this work are in the region near the MS and at a Galactic latitude of GLAT$\leq0\degrees$. To determine the primary association of an absorber, we follow the methodology outlined in Sections 3 and 4, which follows the flowchart presented in Figure \ref{fig:flowchart}. The absorbers are categorized into one of four distinct groups: MS-associated (``MS"), secondary MS-associated (``UMS"), galaxy associated (``G"), and uncertain (``U"). The UMS absorbers are initially classified as uncertain but are assigned secondary association to the MS because of their closer proximity to the MS compared to other nearby galaxies (see Section \ref{sec:determine_assoc} for further details).

\section{Setting up a Model to Associate \HI\ with the Magellanic Stream} \label{sec:distance_model}
\begin{figure}
\centering
\includegraphics[width=\linewidth]{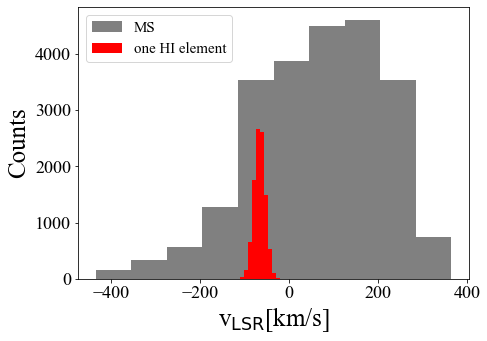}
\caption{The gray histogram shows the velocity distribution of all \HI\ elements in the MS that we adopted from \cite{Nidever08}, while the red histogram shows the empirical Gaussian distribution of an \HI\ element ${\mathcal {N}}(v_{\rm HI}, \sigma_{\rm v, HI}^2)$, where $v_{\rm HI}$ is the velocity centroid and $\sigma_{\rm v, HI}$ indicates the velocity dispersion. See Section \ref{sec:step1} for more details.}
\label{fig:hi_global_demo}
\end{figure}

\begin{figure*}[t]
\centering
\includegraphics[scale=0.55]{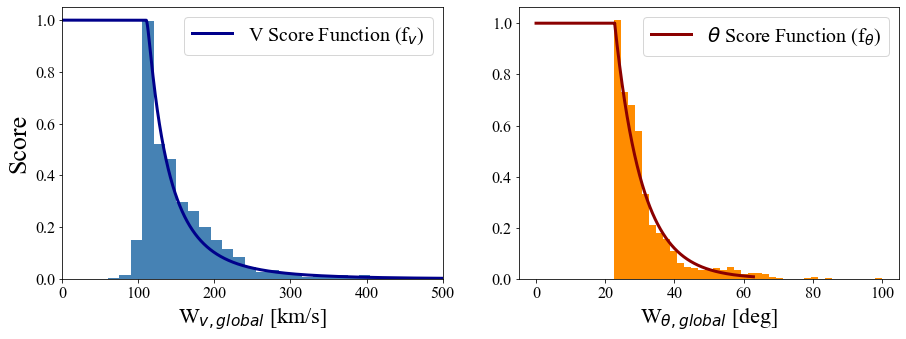}
\caption{Histograms (colored shades) and score functions (solid lines) of the global \HI\ velocity (left) and spatial (right) Wasserstein distances (WDs) between every \HI\ element and the whole MS in the LSR frame (see an example in Figure \ref{fig:hi_global_demo}). With the histograms, we evaluate the best-fit probability distribution functions (PDF) and use them to formulate the score functions ($f_v$ and $f_\theta$). See Section \ref{sec:step1} for more details. }
\label{fig:hi_global_score}
\end{figure*}

\begin{figure}[h!]
    \hspace{-3cm}    
    \includegraphics[width=1.8\linewidth]{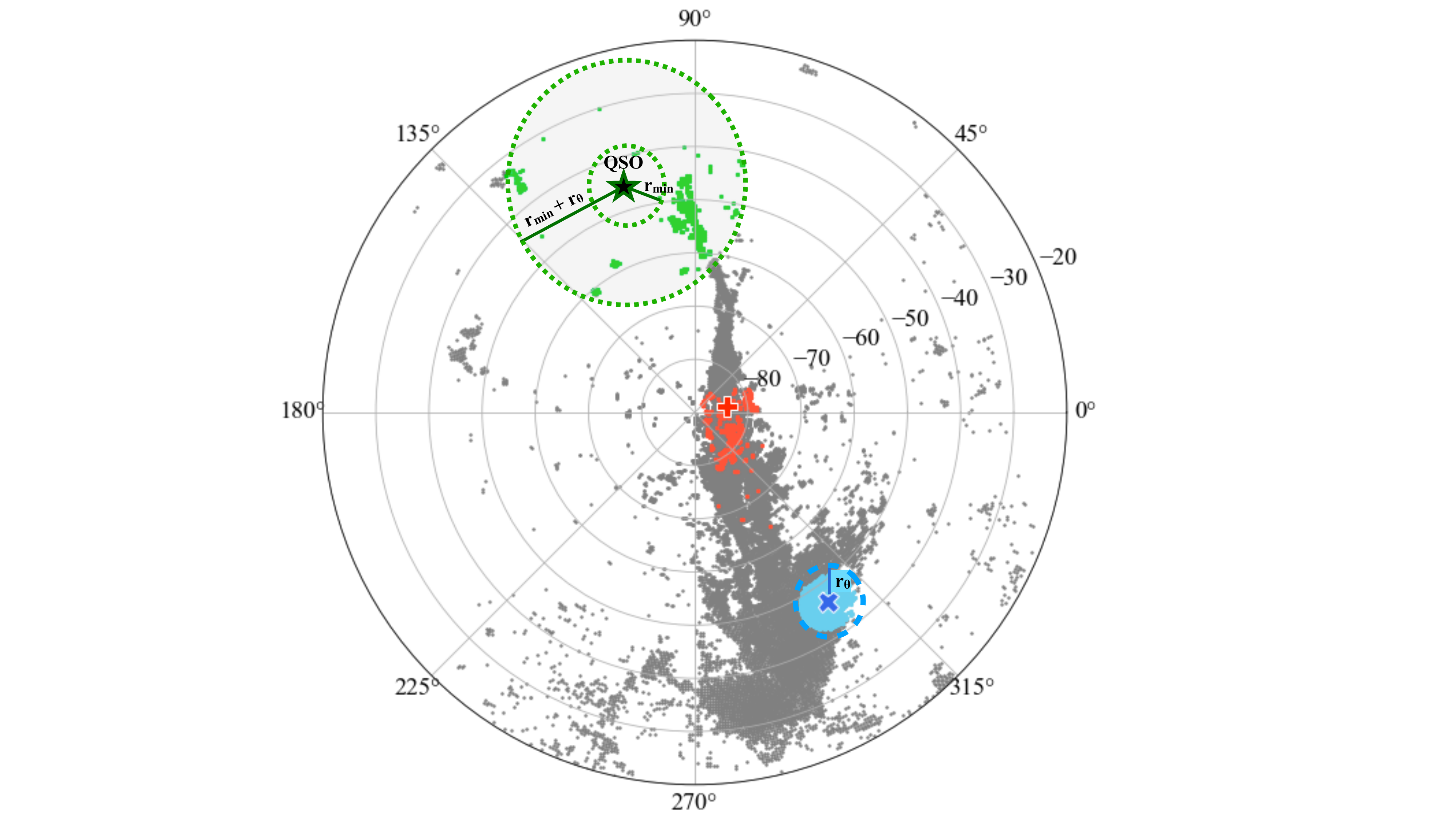}
    \caption{Demonstration of ``local" patches as defined in Section \ref{sec:step2}. The blue ``X" and red ``+" represent locations of example \HI\ elements in the MS. The sky blue pixels near ``X" demonstrate a patch within a radius $\theta$, while the red pixels near ``+" indicate a velocity unit located within $1\sigma$ of the central velocity of the corresponding \HI\ element. The green pixels demonstrate an annulus of two concentric circles, near an absorber shown as the green star.}
    \label{fig:dist_diagram}
\end{figure}

\begin{figure*}[t]
\centering
\includegraphics[width=0.95\linewidth]{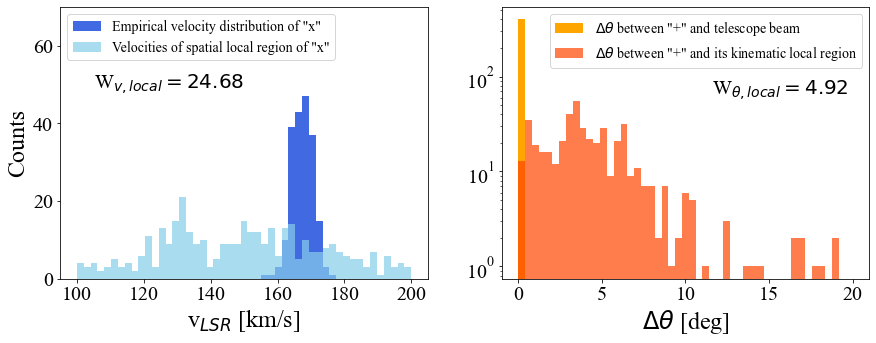}
\caption{Left: Sky blue histogram indicates the velocity distribution of those sky blue pixels highlighted in Figure \ref{fig:dist_diagram}, and the dark blue histogram indicates the modeled velocity distribution of a LOS element marked by the dark blue ``X" in Figure \ref{fig:dist_diagram}. We compute the WD between the two distributions and evaluate the velocity score based on the velocity WD. Right: the red histogram shows the angular separation between those red pixels and the center red ``+" in Figure \ref{fig:dist_diagram}. The orange distribution denotes the beam uncertainty of the LAB data. We compute the WD between the red and orange distributions and use this WD to evaluate the spatial score.}
\label{fig:hi_dist}
\end{figure*}

In this section, we quantify the association likelihood of an \HI\ element to the overall \HI\ extent of the MS based on the \HI\ Gaussian fit results from \citetalias{Nidever08} as described in Section \ref{sec:hi_data}. We elaborate on our methods in the following Sections \ref{sec:step1}--\ref{sec:step2} and Figures \ref{fig:hi_global_demo}--\ref{fig:hi_score_distribution}, and provide a flow chart in Figure \ref{fig:sec3_summary} to summarize all the steps involved.

To quantify the association, we assume the \HI\ within the MS is one cluster\footnote{A cluster is the grouping of objects that share similar features or characteristics.} and evaluate how ``similar" an individual \HI\ element is with the overall \HI\ gas of the MS. We exploit a concept of distance to quantify the degree of ``similarity" amongst every MS \HI\ element. 
A statistical distance measures the separation between two statistical objects, and can be used to measure the separation of two probability distributions, or between an individual sample point and a larger population. A smaller distance between two objects indicates that they share similar statistical properties. For instance, two identical distributions will possess a statistical distance of zero, while two entirely unrelated and uncorrelated distributions will exhibit a substantial separation.

We employ a Wasserstein distance (WD) to measure the statistical distance between an \HI\ element to the main body of the MS. The WD is a type of distance based on probability distributions inspired by the optimal mass transportation problem \citep{opt_transport}. The WD quantifies the minimum effort required to transfer the probability mass from one distribution into another. Among the multitude of available statistical distances, the WD offers flexibility to work with either continuous or discrete distributions and yields interpretable results. In one dimension, the WD is written as
\begin{equation}
{\displaystyle W(\mathbb{P}, \mathbb{Q})= \sum_{i=1}^{n} \left|P_i - Q_i\right|}~~,
\end{equation}
where $\mathbb{P}$ denotes the one dimensional (1D) empirical distribution of a dataset $\mathbf{P}$, and $\mathbb{Q}$ is the 1D empirical distribution of another data set $\mathbf{Q}$. In the context of associating an \HI\ element with the MS, $\mathbb{P}$ represents the 1D distribution of either the velocity or spatial information of an \HI\ element, and $\mathbb{Q}$ is another 1D distribution of the same information, but pertaining to the whole MS.

If two statistical objects are ``close" in terms of statistical distance (resulting in a smaller WD), we consider them to be likely associated. However, it is important to define precisely what constitutes ``closeness," as this is a relative concept. To establish this definition, we derive characteristic geometric distance(s) to determine the boundaries of what we consider to be ``close" to the main body of the MS. 
Only once this is established can we develop a comparison metric that gauges the level of association between individual \HI\ elements and the main \HI\ body of the MS. Developing the distance metric (later called $\Sigma$) to quantify association comprises two distinct procedures, each employing different types of WDs. We first construct a score\footnote{A score in this paper differs from the score used in statistics. We use score as a way to quantify the association.} function to define a relative closeness based on the statistics of a global WD ($W_{\rm{global}}$) in Section \ref{sec:step1}.  We evaluate the association level by applying the score function to a local WD ($W_{\rm{local}}$), then finalize a distance metric $\Sigma$ in Section \ref{sec:step2}.

\subsection{Finding Characteristic Distances of \HI\ in the MS} \label{sec:step1}
In this section, we calculate the WDs that characterize the \HI\ within the MS in both spatial and velocity dimensions. This procedure is analogous to determining the radius of a cluster, if we consider the MS as a single cluster. However, our approach differs slightly from conventional clustering techniques. Typically, a cluster radius serves as a hard cutoff to distinguish between cluster members and non-members. Instead of adopting this hard classification, we develop a score function that provides conditional probabilities and perform classification based on estimated probabilities.

In the velocity dimension, we calculate the WD between the velocity distribution of the entire MS and the empirical velocity distribution of each \HI\ element. In Figure \ref{fig:hi_global_demo}, the velocity distribution of the entire MS is depicted in grey, while the velocity distribution of a single \HI\ element is shown in red. The empirical velocity distribution of an \HI\ element is modeled as a Gaussian distribution, ${\mathcal {N}}(v_{\rm HI}, \sigma_{\rm v, HI}^2)$ with $v_{\rm HI}$ as the centroid velocity and $\sigma_{\rm v, HI}$ as the velocity dispersion. We denote the velocity WD computed between the entire MS and a single \HI\ element as \wglobalv, in a unit of km s$^{-1}$. We repeat this process for all available \HI\ elements and obtain a distribution of all \wglobalv, which is illustrated as a blue histogram in the left panel of Figure \ref{fig:hi_global_score}.

Next, we compute the angular separation between the spatial coordinates of the entire MS and each \HI\ element, and define this spatial WD as \wglobalx\ in unit of degrees. We begin by computing the angular separation between an individual \HI\ element and the entire MS and subsequently calculate the WD between this angular separation and the positional uncertainty induced by the LAB data. This uncertainty due to the beam is modeled as a Gaussian distribution, ${\mathcal {N}}(0, \theta_{\rm{beam}})$, with $\theta_{\rm{beam}}$($\approx 0.6\degrees$) representing the full width at half maximum (FWHM) of the beam. The effect of the beam is almost negligible when contrasted with the angular separation between an \HI\ element and the entire MS. We repeat this process for all available \HI\ elements within the MS and generate a distribution of all \wglobalx, which is illustrated as an orange histogram in the right panel of Figure \ref{fig:hi_global_score}.

The ``global" WD (\wglobalv\ and \wglobalx) distributions, presented in Figure \ref{fig:hi_global_score} provide an overview of how the \HI\ elements are distributed in velocity and spatial dimensions. The peaks in the \wglobalv\ and \wglobalx\ distributions signify relative clustering distances among the \HI\ elements. Higher values of \wglobalv\ and \wglobalx\ (moving further away from the peaks) indicate that a particular \HI\ element is more distant from the majority of \HI\ in the MS. As shown in Figure \ref{fig:hi_global_score}, the majority of \HI\ elements are located near the peaks of the distributions, and decrease at larger \wglobalv\ and \wglobalx\ values beyond the peaks. We leverage the decreasing trends in the ``global" WDs to create the score functions.

To ensure the continuity and completeness of the scores, we fit probability distribution functions (PDFs) to both distributions in Figure \ref{fig:hi_global_score}. We choose the best-fit PDF for each distribution as the one that has the highest score in the Kolmogorov-Smirnov goodness-of-fit test, which is conducted between the histogram distribution and the PDFs available in the \texttt{scipy} package. 
We find that the PDF of 
the \wglobalv\ distribution can be best fit by a Birnbaum-Saunders distribution function expressed as 
\begin{equation*}
g(x,c) = \frac{x+1}{2c\sqrt{2\pi x^3}} {\rm exp}(-\frac{(x-1)^2}{2xc^2}),
\end{equation*}
where $x$ is a variable greater than 0 and $c$ is a shape parameter greater than 0. And the distribution of \wglobalx\ can be best represented by another PDF, Kappa 3 distribution expressed as
\begin{equation*}
h(x, \alpha) = \alpha(\alpha+x^\alpha)^{(-(\alpha+1)/\alpha)},  
\end{equation*}
where $\alpha$ is a shape parameter.

The smaller the WD, the more likely an \HI\ element is associated with the MS. Therefore, the scores should be the highest for relatively small WD values and gradually decrease as \wglobalv\ and \wglobalx\ increase. We define the most densely populated \wglobalv\ and \wglobalx\ values as $\omega_{\rm{v}, global}$ (=134.6 $\kms$) and $\omega_{\rm{\theta}, global}$ (=22.7$\degrees$), and assign the highest scores of unity (1) at these values (near the peaks in Figure \ref{fig:hi_global_score}).
We employ the best-fit PDFs to transform the remaining global WD values into normalized scores. These modified best-fit PDFs become the score functions ($f_v$ and $f_\theta$), which are represented as solid lines in Figure \ref{fig:hi_global_score}. More specifically, we define the velocity score function as
\begin{equation}
f_v(W_{v}) =
    \begin{cases}
    1, & \text{if } W_{v}\leq \omega_{\rm{v}, global}\\
    g(W_{v})/\max(g(W_{v})),& \text{otherwise}
\end{cases}
\label{eq:vscore}
\end{equation}
where $W_{v}$ is any velocity WD and $g(W_{v})$ is the Birnbaum-Saunders distribution function evaluated at a given $W_{v}$. We define the spatial score function as
\begin{equation}
f_{\theta}(W_{\theta}) =
    \begin{cases}
    1,& \text{if } W_{\theta}\leq \omega_{\rm{\theta}, global} \\
    h(W_{\theta})/\max(h(W_{\theta})),& \text{otherwise}
\end{cases}
\label{eq:xscore}
\end{equation}
where $W_{\theta}$ is any spatial WD and $h(W_{\theta})$ is the Kappa 3 distribution function evaluated at a given $W_{\theta}$.

Thus far, we have constructed the score functions by calculating the WDs between individual \HI\ elements and the entire MS. The score functions quantify the relative level of association of each \HI\ element with the MS. In the subsequent section, we evaluate the scores of all \HI\ elements in the MS in an unbiased way to further refine the characteristic WDs and construct the distance metric $\Sigma$ to evaluate the level of association near the MS.

\subsection{Constructing a Distance Metric $\Sigma$} \label{sec:step2}
The \HI\ score functions are designed to quantify \HI\ associations. For instance, we can quantify the association in velocity space between one \HI\ element and the whole \HI\ extent of the MS based on a measured velocity WD, $W_{v}$, and a normalized velocity score $f_v(W_{v})$, where a higher score indicating higher association. However, we do not directly assign scores based on individual \wglobalv\ and \wglobalx\ values that we show in Figure 3. The global WDs measure the WD between an \HI\ element and all \HI\ emission in the MS. If we were to naively determine the level of association solely based on scores derived from \wglobalv\ or \wglobalx, elements in the regions with the highest \HI\ covering factor would consistently score the highest, while elements in the regions with the lowest \HI\ covering factor would score the lowest in association. This is due to the morphological asymmetry and uneven distribution of \HI\ emission within the MS, as illustrated in Figure \ref{fig:MS_full}. A high coverage may be a sufficient, yet not necessary, condition for the association.

To mitigate these \HI\ covering factor biases, we only build the score functions upon the global WDs (\wglobalv\ and \wglobalx; Section \ref{sec:step1}) and separately derive two local WDs (hereafter \wlocalv\ and \wlocalx) when quantifying the association of an \HI\ element. The local WD minimizes the biases due to covering factor variation across the MS while conserving the general characteristics of the MS. This approach not only prevents the bias introduced by the uneven distribution of \HI\ in the MS but also provides robust constraints on the scoring metric. Any structures, including gaseous media, are more likely to be strongly coupled kinematically and spatially at a short range. Consequently, a distance measured between an object and its kinematically and spatially ``local" areas is likely to be smaller than a distance measured between the same object and the entire gaseous structure to which it belongs, even if the representation is unbiased. By determining the level of association in a local area using the score function derived from the overall structure, our model becomes more resilient to data outliers.

In Figure \ref{fig:dist_diagram}, we illustrate our definition of a spatial ``local" region for an \HI\ element. The \HI\ element is shown as a blue ``X", and the sky blue pixels around the ``X" demonstrate a spatial local region, defined as an annulus of two concentric circles, both centered at the location of the \HI\ element ``X". The inner concentric circle has a radius of $r_{\rm min}$ that indicates the minimum angular distance to the nearest \HI\ element from the \HI\ element ``X"; because of the prevalent presence of \HI, $r_{\rm min}$ is always 0 for any \HI\ element. Note that this is different from the $r_{\rm min}$ values set for UV absorbers (green pixels) that we discuss in Section \ref{sec:absorber_hi}. 

The outer concentric circle has a radius of $r_{\rm min} + r_{\theta}$, where $r_{\theta}$ is an arbitrary radius that captures a sufficient number of \HI\ elements (sky blue pixels) to preserve the characteristics of the MS. Naturally, the spatial local region of an \HI\ element, such as ``X," becomes the area enclosed by the circle with the radius $r_{\theta}$, which is the sky blue patch surrounding ``X." We initially set $r_{\theta}$ to 2 degrees to capture a 
sufficient number of \HI\ elements at most locations within the MS. We increase $r_{\theta}$ when at least 1$\%$ of the total \HI\ area of the MS is not collected, but keep the maximum $r_{\theta}$ at 7$\degrees$. Beyond 7$\degrees$, we observe a significant increase in the velocity variance within the collected \HI\ patch.

 \begin{figure}[h!]
    \centering
    \includegraphics[width=\linewidth]{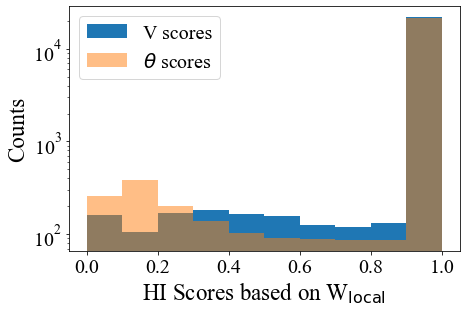}
    \caption{Histogram distributions of velocity (blue) and spatial (orange) scores for all \HI\ elements in the MS. We evaluate the scores by applying the score function derived from Section \ref{sec:step1} to the local WDs (\wlocalx\ and \wlocalv) evaluated for individual \HI\ elements in Section \ref{sec:step2}.}
    \label{fig:hi_score_distribution}
\end{figure}

We further demonstrate a kinematic local region shown as a red ``+" and surrounded by a patch of red pixels in Figure \ref{fig:dist_diagram}. A kinematic local region of an \HI\ element covers all neighboring \HI\ elements that have velocities within the velocity range of the given \HI\ element $v_{\rm HI} \pm \sigma_{\rm v, HI}$, where $v_{\rm HI}$ and $\sigma_{\rm v, HI}$ indicate the central velocity and velocity dispersion of a chosen \HI\ element, respectively. We only consider \HI\ elements within a radius of 40 degrees of the given \HI\ element to avoid selecting unassociated \HI\ elements that happen to have similar velocities but are spatially at large distances. The search radius of 40 degrees 
corresponds to a \wglobalx\ value with a normalized score of 0.1, or $f_\theta^{-1}(0.1)$, as shown in the right panel of Figure \ref{fig:hi_global_score}.

After defining the spatial and kinematic local regions for each \HI\ element, we compute the statistical distances, denoted as local WDs \wlocalx\ and \wlocalv, between the \HI\ element and its local regions. We compute the local velocity WD (\wlocalv) by measuring the WD between the empirical velocity distribution of one \HI\ element and the velocity distribution of its spatial local region. This is similar to the two histogram distributions shown in Figure \ref{fig:hi_global_demo}, but with the gray histogram representing the velocity distribution of a local patch instead of the whole \HI\ extent of the MS. Similarly, we evaluate the local spatial WD (\wlocalx) by measuring the angular separation between one \HI\ element and its kinematic local region. The left panel of Figure \ref{fig:hi_dist} demonstrates the empirical velocity distribution of the ``X" element in dark blue that we model as a Gaussian model ${\mathcal {N}}(v_{\rm HI}, \sigma_{\rm v, HI}^2)$ (same as in Figure \ref{fig:hi_global_demo}), and the velocity distribution of its spatial local region (the sky blue patch in Figure \ref{fig:dist_diagram}) is shown in sky blue. The computed local WDs are indicated in the corresponding panels. The \wlocalv\ between the empirical velocity distribution of ``X" (dark blue distribution) and the velocities of its spatial local region (sky blue distribution) is 24.65 km/s. The right panel of Figure \ref{fig:hi_dist} displays the angular separation ($\Delta \theta$) between the ``+" element and its kinematic local region in the red distribution, along with the angular separation between ``+" and the positional uncertainty induced by the telescope beam in orange. The \wlocalx\ between these two distributions is 4.92 degrees.

After calculating the \wlocalv\ and \wlocalx\ for all \HI\ elements in the MS, we determine the association scores for these local WDs using the score functions (Equations \ref{eq:vscore} and \ref{eq:xscore}) developed in the previous section. Figure \ref{fig:hi_score_distribution} presents the distributions of the velocity scores $f_v$(\wlocalv) in blue and the spatial scores $f_\theta$(\wlocalx) in orange. The normalized scores in velocity and $\theta$ allow us to evaluate the association of \HI\ elements in both velocity and position space simultaneously, which is not possible when comparing the global or local WD values. A vast majority of the \HI\ elements score one, indicating that most \HI\ elements' local WDs are smaller than the most populated global WDs. This is expected because local regions often have more directly associated \HI\ emission. Additionally, we observe a clear break in both score distributions at the 0.9 mark. Since this trend is evident for both scores, we consider the WD values at the score 0.9, $f_v^{-1}(0.9)$ and $f_\theta^{-1}(0.9)$ (117.7 km/s and 25 degrees, respectively) as the characteristic velocity and angular size of the \HI\ in the MS. 

For clarity, we define the local WDs corresponding to the score 0.9 as $\omega_{\rm{v}, local}$ and $\omega_{\rm{\theta}, local}$.   
These characteristic distances also represent the size of the MS as a cluster.

\begin{figure}
    \centering
    \includegraphics[width=\linewidth]{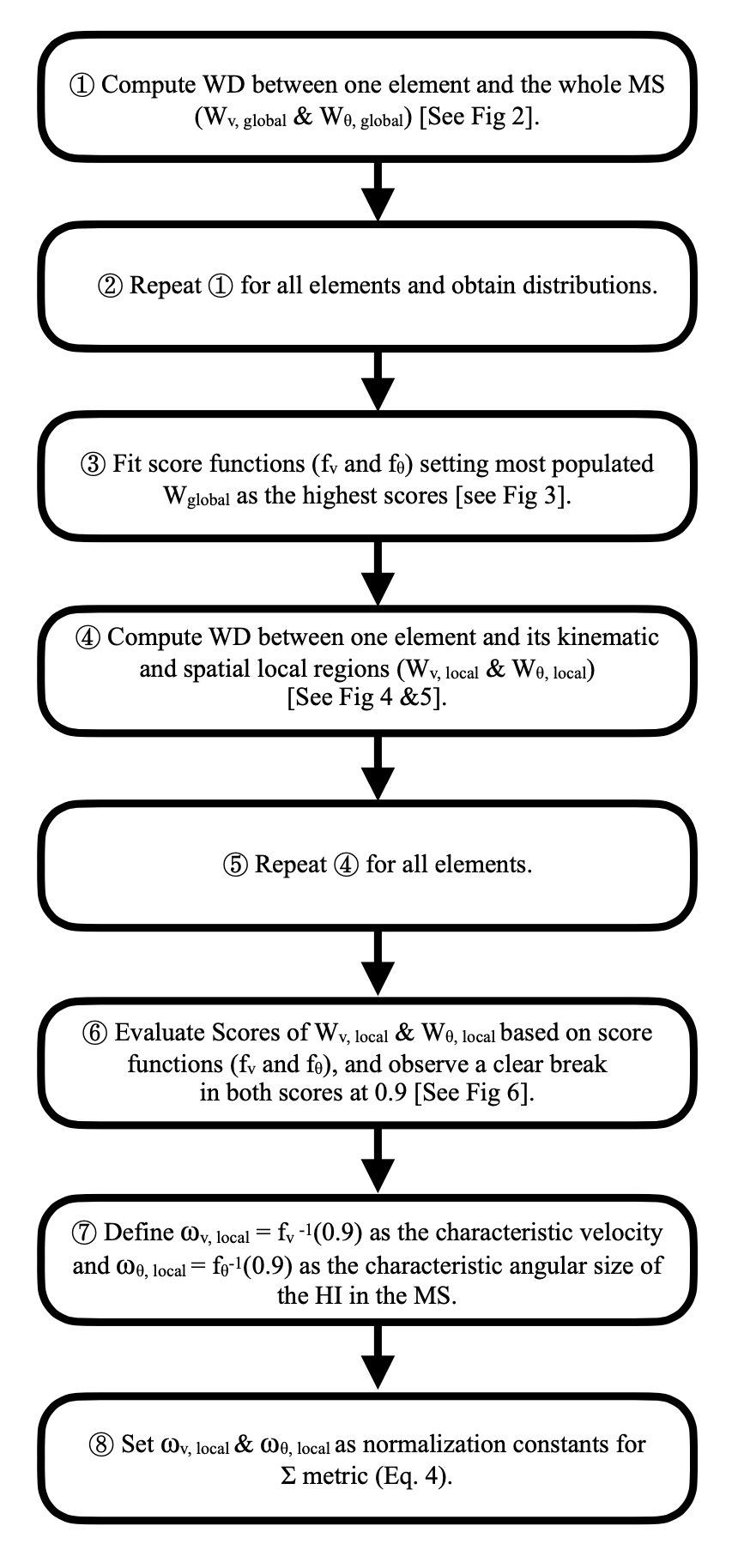}
    \caption{A summary of the steps we implemented in Sections \ref{sec:step1} and \ref{sec:step2} to derive the characteristic distances of the \HI\ in the MS and the $\Sigma$ metric.}
    \label{fig:sec3_summary}
\end{figure}

We define a generic, dimensionless distance metric $\Sigma$ as: 
\begin{equation}
\Sigma = \sqrt{ \left( \frac{W_{v, \rm{local}}}{\omega_{\rm{v}, local}} 
 \right)^2+\left( \frac{W_{\theta, \rm{local}}}{\omega_{\rm{\theta}, local}} \right)^2},
\label{eq:sigma}
\end{equation}
where the characteristic distances ($\omega_{\rm{v}, local}$ and $\omega_{\rm{\theta}, local}$) now serve as normalization constants. Hereafter, we will refer to the distance metric used to evaluate the association of an \HI\ element to its local environment as $\Sigma_{\rm HI}$; this is 
to distinguish the distance metric $\Sigma$ that we develop in Sections \ref{sec:absorber_hi}--\ref{sec:absorber_gal} for UV ion absorbers. We only consider an \HI\ element to be associated with the MS if it meets the threshold of $\Sigma_{\rm HI}\leq\sqrt{2}$, 
and lower $\Sigma_{\rm HI}$ values indicate higher levels of associations. 
The threshold of $\Sigma_{\rm HI}\leq\sqrt{2}$ is set such that the \wlocalv\ and \wlocalx\ are at the MS characteristic distances of 117.7 km/s and 25 degrees, respectively. We summarize all the steps to derive the characteristic distances of the \HI\ in the MS and the $\Sigma$ metric in Figure \ref{fig:sec3_summary}.

\begin{figure*}[t]
\centering
\includegraphics[width=\linewidth]{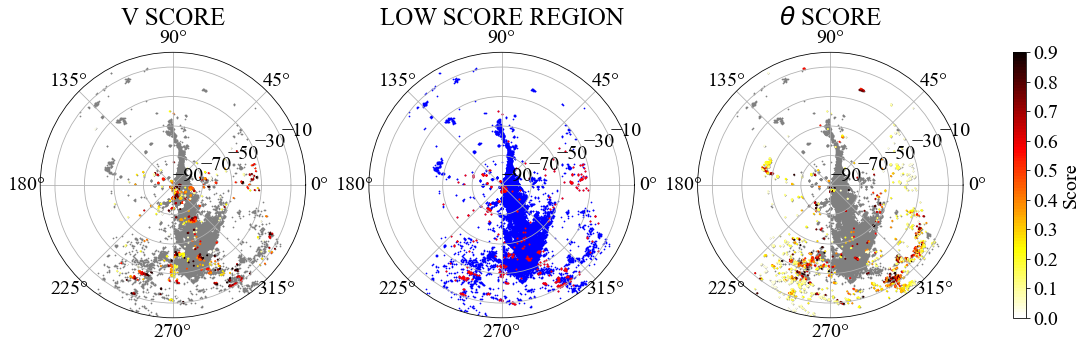}
\caption{The score distributions of the \HI\ in the MS. On the left, we color-code the \HI\ elements with $f_v \leq 0.9$, which we consider to be less likely to be associated with the MS. On the right, we color-code those with spatial scores below 0.9. In the center, we show in red the \HI\ elements with both velocity and spatial scores less than 0.9 and have $\Sigma$ greater than $\sqrt{2}$; these \HI\ elements are considered to be not associated with the MS and are not used in the subsequent absorber analysis.}
\label{fig:hi_score_distr}
\end{figure*}

\subsection{Estimating Model Uncertainties}\label{sec:error}
We estimate the \HI\ score PDF uncertainties using bootstrapping \citep{bootstrap}. Because all score metrics are evaluated based on the \HI\ data from \citetalias{Nidever08}, we apply bootstrapping by randomly discarding 20 $\%$ of all the \HI\ data, then draw at random a sample with replacement until the number of new draw in each bootstrap realization is equal to the total number of the \HI\ elements in the original dataset. For each realization, we fit PDFs to the sampled velocity and angular separation WD distributions, then re-evaluate the scores based on the bootstrapped PDFs. The mean and variance of the velocity and spatial scores are estimated from 1000 bootstrap realizations, and are found to be insensitive to increasing or decreasing numbers of bootstrap realizations.

\subsection{Results on \HI\ Associations}
\label{sec:hi_association_result}
With the \HI\ distance metric $\Sigma_{\rm HI}$ and local WDs evaluated in Sections \ref{sec:step1} and \ref{sec:step2}, we now quantify the level of associations of individual \HI\ elements to the main body of the MS. In Figure \ref{fig:hi_score_distr}, we show the velocity scores $f_v$(\wlocalv) in the left panel and the spatial scores $f_\theta$(\wlocalx) in the right panel for all \HI\ elements available in \citetalias{Nidever08}'s dataset.

An \HI\ velocity score (on the left panel) indicates the kinematic proximity between an \HI\ element and its spatially local region. It measures the extent to which the velocity characteristics of an \HI\ element aligns with the overall velocity distribution of the MS. Lower velocity scores mean weaker associations, which suggest the presence of complex dynamics or interactions in local areas. Notably, regions with low velocity scores are often associated with specific features, such as the intersection of two \HI\ flows near the south Galactic pole (located at the center of the polar map) as previously noted by \cite{putman2003}. Additionally, some areas with low-velocity scores are found in close proximity to the Large and Small Magellanic Clouds (LMC and SMC) near the bottom of the polar-projected map.

A spatial score (on the right panel) reveals the spatial proximity between an \HI\ element and its kinematic local region. It offers insights into the relative spatial scales of kinematically homogeneous \HI\ elements. The main body of the MS displays a transverse velocity gradient from the head (near the LMC and SMC) to the tail (near M31; see Figure \ref{fig:MS_full}). Naturally, segmented clouds situated distant from the main body tend to have low spatial scores because their kinematic characteristics deviate from the prevailing trend, such as the clouds located near the head of the MS with Galactic latitudes between $-30\degrees$ and $0\degrees$ in the right panel of Figure \ref{fig:hi_score_distr} (lower section of the polar map). Interestingly, \HI\ elements near the tail region of the MS exhibit a higher degree of kinematic consistency with uniformly high spatial scores, even though the area is known to be more fragmented.

In the middle panel of Figure \ref{fig:hi_score_distr}, the \HI\ elements with low combined scores are depicted in red, while the rest are shown in blue. We define \HI\ elements with low combined scores as those with both velocity and spatial scores below 0.9 and with $\Sigma_{\rm{HI}}$ values greater than $\sqrt{2}$. The \HI\ elements with low combined scores are found in regions with inhomogeneous spatial and kinematic distributions that are indicative of increased complexity, such as the intersection of colliding \HI\ flows, in proximity to the LMC and SMC, or within fragmented clouds with greater kinematic complexity compared to the MS.

\section{Identifying Absorber Associations} \label{sec:assoc_eval} 
In Section \ref{sec:absorber_hi}, we move on to use the distance metric $\Sigma$ (Eq.~\ref{eq:sigma}) to quantify the associations between UV absorbers and the \HI\ within the MS, which we refer to as $\Sigma_{\rm MS}$. In Section \ref{sec:absorber_gal}, we quantify the associations between UV absorbers and halos of nearby galaxies in close projection as $\Sigma_{\rm G}$. In Section \ref{sec:determine_assoc}, we compare the $\Sigma_{\rm{MS}}$ and $\Sigma_{\rm{G}}$ values to determine the dominant association of a given absorber. Lastly, we compare our algorithm with the D parameter method developed by \cite{Peek_Dparam} in Appendix \ref{sec:peek_compare}.

\subsection{Absorber Associations with the \HI\ MS} \label{sec:absorber_hi}
When assessing the association between an absorber and its local \HI\ environment, we follow a procedure that closely resembles the one outlined in Section \ref{sec:step2}. We first filter out \HI\ elements in the \citetalias{Nidever08} data that are unlikely to be associated with the MS, which is quantified as those with both spatial and velocity scores less than 0.9 and $\Sigma_{\rm HI}$ greater than $\sqrt{2}$ (red pixels in the middle panel of Figure \ref{fig:hi_score_distr}; see Section \ref{sec:hi_association_result} for further details). Hereafter, the ``main body of the MS" or ``\HI\ extent of the MS"  is referred to those \HI\ elements that our algorithm identifies to be associated with the MS, shown as blue pixels in the middle panel of Figure \ref{fig:hi_score_distr}.

We define a spatial local \HI\ region of a UV absorber as the set of \HI\ elements enclosed within an annulus formed by two concentric circles; we show an example of a spatially local \HI\ region for an absorber in green pixels in Figure \ref{fig:dist_diagram}. Unlike the steps outlined for \HI\ associations in Section \ref{sec:step2}, $r_{\rm min}$ for UV absorbers are not necessarily zero because most absorbers are located beyond the \HI\ extent of the MS. While we initiate with $r_{\theta} = 2\degrees$, we do not enforce a maximum limit at 7 degrees as we did in Section \ref{sec:step2}. Instead, we progressively increase $r_{\theta}$ until we collect at least 1$\%$, the minimum flux which captures local properties without imposing any biases, of the total \HI\ area in the MS.

We identify a kinematic local \HI\ region of a UV absorber as the area within which the \HI\ elements all have velocities within the velocity range defined by the absorber's minimum and maximum velocities that we tabulate in Table \ref{tb:qso}. We collect at least 5\% of the \HI\ elements in the MS, which is more than the percentage (1\%) collected for spatial local \HI\ regions to better reflect the MS's local morphology. Again, unlike the case of \HI, we do not set a specific maximum limit in area sizes.

We use the same method as described in Section \ref{sec:step2} to compute the spatial WD between a UV absorber and its kinematic local \HI\ region, and the velocity WD bewteen the absorber and its spatial local \HI\ region. 
The key difference from Section \ref{sec:step2} is the modeling of the absorber's empirical velocity distribution depending on whether the absorber's data are calculated based on the AOD or the VP methods. For absorbers originating from VP data sets (e.g., \citetalias{fox20} and \citetalias{Krishnarao22}), we model the empirical velocity distributions as ${\mathcal {N}}(v_{\rm ion},\sigma_{\rm v, ion})$, where $v_{\rm ion}$ is the fitted centroid velocity of an absorber in the local standard of rest (LSR) frame, and $\sigma_{\rm v, ion}$ is the velocity dispersion. For absorbers from the AOD data sets (e.g., \citetalias{richter17}, \citetalias{lehner20_amiga}, and our archive sample), we model the velocity distributions as continuous uniform distributions bound by the absorbers' minimum and maximum velocities, i.e., $\mathcal{U}(v_{\rm{min}}, v_{\rm{max}})$ for each ion. We denote the velocity WD between an absorber and its spatial local \HI\ region as W$_{\rm v, ion}$, and the spatial WD between an absorber and its kinematic local \HI\ region as W$_{\rm \theta, ion}$. The distance metric to quantify the associations between UV absorbers and the main body of the MS, $\Sigma_{\rm MS}$, is calculated based on the absorbers' spatial and velocity WDs (W$_{\rm v, ion}$, W$_{\rm \theta, ion}$) using Eq.\ref{eq:sigma}.

\subsection{Possible Associations with Galaxies} \label{sec:absorber_gal}
An additional source of absorption, apart from the MS, is metals in the CGM of nearby galaxies that are further than the MS in distances but appear to be in close projection. In this section, we use the distance metric $\Sigma$ to to evaluate possible associations between our UV absorbers and nearby galaxies. We refer to the $\Sigma$ calculated between an absorber and the modeled dark matter halo of a galaxy as $\Sigma_{\rm{G}}$.

To determine the association of an absorber with a galaxy, we first need to quantify a galaxy's physical influence by computing its virial radius and escape velocity from the galaxy's stellar mass. 

\begin{figure}[h!]
    \hspace{-0.7cm}
    \includegraphics[width=1.12\linewidth]{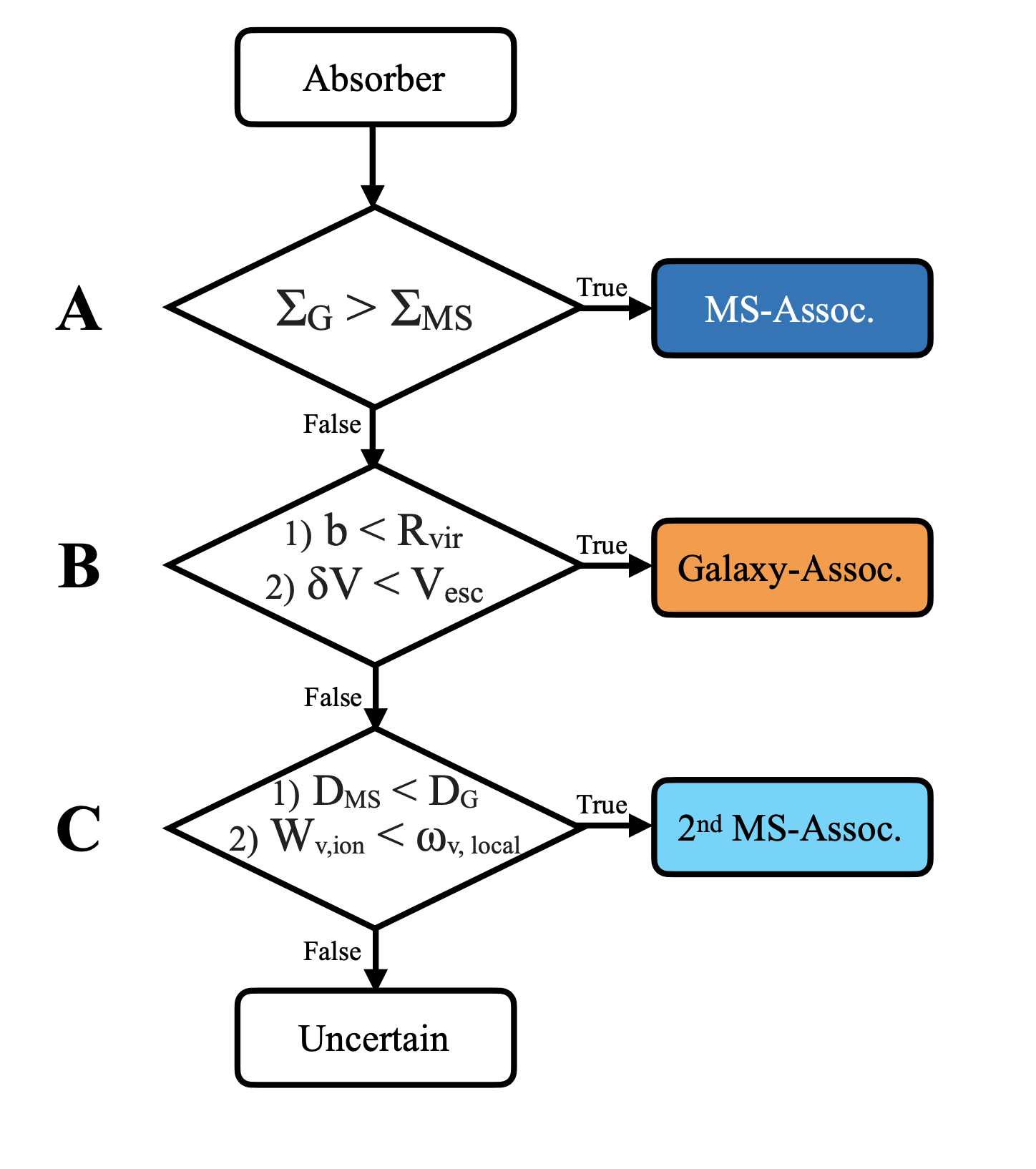}
    \caption{A flow chart of how the dominant association of an absorber is determined, which is described in detail in Section 4.3. Overall, we implement three criteria (A, B, and C) to determine an absorber's association based on its distance metric ($\Sigma_{\rm MS}$, $\Sigma_{\rm G}$) with respect to the MS and nearby galaxies (Criterion A), as well as its proximity to nearby structures in physical position and velocity space (Criteria B or C).}
    \label{fig:flowchart}
\end{figure}

We estimate the galaxy's stellar mass by employing the K-band luminosity ($K_L$), which is an indicator of the stellar mass of a galaxy due to its insensitivity to internal extinction and young stellar populations \citep{bell_kband_lumin}. We adopt $K_L$ values from \cite{Karachentsev13} and convert them into stellar masses using the unity relation \citep{bell_kband_lumin}. We compute the halo masses using a stellar-to-halo mass relation from \cite{Behroozi13_smhm}, and define the halo masses based on two density contrast definitions: one relative to the mean matter density of the Universe ($\rho_{\rm{m}}$) and another relative to the critical density ($\rho_{\rm{c}}$) \citep{davis_halo_density}. We consider both density definitions to provide a more conservative assessment of an absorber's association with a galaxy. Using the virial theorem, we derive two types of virial radii ($R_{200m}\And R_{200c}$ $\in R_{\rm vir}$) and their respective escape velocities ($V_{200m}\And V_{200c}$ $\in V_{\rm esc}$) for a galaxy. For galaxies without $K_L$ values, we directly adopt the halo masses and virial radii from \citetalias{putman21} and \citetalias{lehner20_amiga}.

Next, we evaluate the spatial and velocity WDs based on the two sets of virial radii and escape velocities. The velocity WD is computed between the empirical velocity distribution of an absorber (as discussed in Section \ref{sec:absorber_hi}) and the velocity distribution of a galaxy ${\mathcal {N}}(V_{\rm{G}}, V_{\rm{esc}})$, where $V_{\rm{G}}$ represents the radial velocity of the galaxy in the LSR frame and $V_{\rm{esc}}$ denotes the galaxy's escape velocity. 

Before computing the spatial WD, we model the angular distribution of the halo of a galaxy, which is computed as the area enclosed within the virial radius ($R_{200m}$ or $R_{200c}$) projected to the distance to the galaxy. We then evaluate the spatial WD between the angular separation of an absorber and the effective angular distribution of the galaxy's halo. We note that the positional uncertainty due to the size of the COS aperture (2.5 arcsecond) is negligible.

\subsection{Determining Dominant Associations} \label{sec:determine_assoc}
So far, we have calculated $\Sigma$ values relative to the MS ($\Sigma_{\rm{MS}}$) and nearby galaxies ($\Sigma_{\rm{G}}$). In this section, we compare the $\Sigma$ values to identify the dominant association of each absorber. 

Figure \ref{fig:flowchart} provides an overview of the process we use to determine the dominant association of an absorber. We identify an absorber as MS-associated when $\Sigma_{\rm{MS}} < \Sigma_{\rm{G}}$ for all nearby galaxies (Criterion A in Figure \ref{fig:flowchart}). To be conservative in identifying galaxy associations, we recognize an absorber as galaxy-associated only when  $\Sigma_{\rm{MS}} > \Sigma_{\rm{G}}$ and \textbf{all} of the following three conditions are met (Criterion B):

\begin{enumerate}
    \item An absorber's impact parameter $b$ is less than the virial radius of a galaxy: $b<R_{\rm vir}$.
    \item The absolute difference between the velocity centroid of an absorber and the velocity of a galaxy is less than the escape velocity of the galaxy: $\delta V\equiv \lvert v_{\rm ion} - V_{G} \rvert < V_{esc}$.
    \item The above conditions are satisfied for the pairs of virial radius and escape velocity values derived based on both $\rho_{m}$ and $\rho_{c}$ as we describe in Section \ref{sec:absorber_gal}.
\end{enumerate}

When multiple galaxies are found with $\Sigma_{\rm{MS}} > \Sigma_{\rm{G}}$ and Criterion B is met, we consider the absorber to be primarily associated with the galaxy that has the smallest $\Sigma_{\rm{G}}$, velocity offset $\delta V$, and impact parameter $b$. However, if all these galaxies show similar $\Sigma_{\rm{G}}$, $\delta V$, and $b$ values such that a primary association cannot be determined, we recommend all possible galaxy associations for the absorber. 

\begin{figure*}[t]
\includegraphics[width=\linewidth]{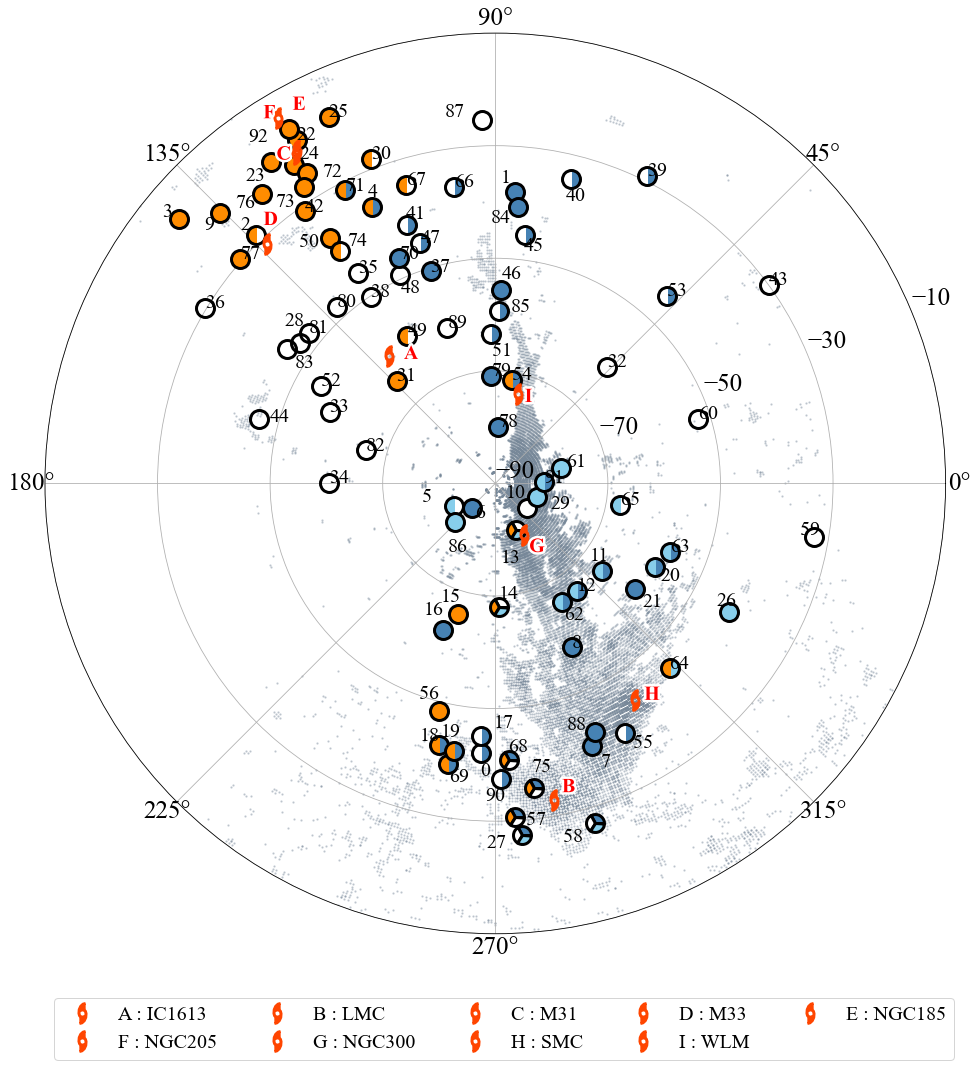}
\caption{The projected map of the absorbers (in numbers), galaxies (in alphabet letters), and the \HI\ (gray background). The colors indicate the identified association: blue for the MS-associated, light blue for the UMS (secondary MS associated), orange for the galaxy-associated, and white for uncertain. The UMS indicates a group of absorbers assigned the secondary association to the MS because they only pass Criterion C in Figure \ref{fig:flowchart}, but not Criterion A. Circles with more than one colored wedge have multiple absorbers detected along the same sight lines at different velocities, and the absorbers are found to have different associations by our algorithm. We show all galaxies associated with at least one absorber with a red galaxy marker and red letter, which follows the same notation as in Table \ref{tb:gal_info}.}
\label{fig:qso_assoc_diagram}
\end{figure*}

\startlongtable
\begin{deluxetable*}{cccccccccc}
\tabletypesize{\footnotesize}
\tablecaption{List of Associated Galaxies}
\tablehead{\colhead{Label}\vspace{-0.2cm}  & \colhead{Name} & \colhead{GLON} & \colhead{GLAT} &  \colhead{V$_{\rm{LSR}}$}  & \colhead{Distance} & \colhead{log M$_{\rm{HI}}$}  & \colhead{log M$_{\rm{halo}}$} & \colhead{R$_{200m}$} & \colhead{R$_{200c}$}\\ \\[-0.45cm]
\colhead{} & \colhead{} & \colhead{[deg]} & \colhead{[deg]} & \colhead{[km/s]} & \colhead{[Mpc]} & \colhead{log [M$_{\odot}$]} & \colhead{log [M$_{\odot}$]}  & \colhead{[kpc]} & \colhead{[kpc]}\\ 
\colhead{(1)} & \colhead{(2)} & \colhead{(3)} & \colhead{(4)} & \colhead{(5)}  & \colhead{(6)} & \colhead{(7)} & \colhead{(8)} &\colhead{(9)} & \colhead{(10)}} 

\startdata 
A	& IC1613 &	129.73 &	$-$60.56	&$-$236.39	&0.76&	7.80 & 10.59 & 106 & 71 \\
B	&	LMC	&280.47&	$-$32.89	&265.66&0.05	&8.66	& 11.36	& 191 & 129 \\
C	&	M31&	121.17 &	$-$21.57&	$-$293.36	&0.77	&9.73&	12.43 & 435 & 293 \\
D	&	M33	&133.61	&$-$31.33&	$-$183.52 &	0.93&	9.40	&11.42	& 200 & 135 \\
E	&	NGC185 &	120.79&	$-$14.48	&$-$199.23&	0.66	&5.08 & 10.75 & 120 & 81 \\
F	&	NGC205	&120.72	&$-$21.14&	$-$218.19	&0.80&	5.58 &	11.13 &	160 & 108 \\
G & NGC300 & 299.21&$-$79.42&	138.0&	2.09&	9.32&	11.37 & 193 & 130 \\
H	&	SMC	& 302.81	&$-$44.33	& 149.38 & 0.06 &	8.65&	11.10 & 157 & 106 \\
I & WLM & 75.86 & $-$73.62 & $$-$$124.82 & 0.98 & 7.84 & 10.31 & 85 & 58 \\
\enddata
\smallskip
\tablenotetext{}{Note: Col (1): Alphabet letters assigned to absorber associated galaxies. 
Col (2): Galaxy names. 
Cols (3) \& (4): Locations of galaxies in Galactic coordinates. 
Col (5): Velocities in the LSR frame.
Col (6): Distances from \cite{Karachentsev13}.
Col (7): The \HI\ Masses from \cite{Karachentsev13}.
Col (8): Halo masses of galaxies, assuming a unity relation between K-band luminosities from \cite{Karachentsev13} and stellar masses, which are then converted to halo masses based on the stellar mass-halo mass relation from \cite{Behroozi13_smhm}.
Col (9):  Virial radii, defined as the radii within which the mean density is 200 times the mean matter density at $z=0$.
Col (10):  Virial radii, defined as the radii within which the mean density is 200 times the critical density at $z=0$.} 
\label{tb:gal_info}
\end{deluxetable*}

Since we cross-examine between two virial radius and escape velocity definitions and impose rather strict conditions in Criterion B, the numbers of galaxy-associated absorbers and its associated galaxies are relatively small, as shown in Table \ref{tb:gal_info}. Many absorbers that may well be associated with either the MS or nearby galaxies may not have any associations assigned by our algorithm. 
For such absorbers, we implement Criterion C to consider the physical distance between an absorber and the MS's \HI\ content or nearby galaxies. Assuming a distance to the MS of 100 kpc \citep{besla12}, we compare the physical distance ($D_{\rm{MS}}$) between an absorber and its spatial local \HI\ region, and the physical distance ($D_{\rm{G}}$) between the absorber and all galaxies that yield $\Sigma_{\rm{G}} < \Sigma_{\rm{MS}}$. If an absorber's $D_{\rm{MS}}$ is smaller than all the $D_{\rm{G}}$ values and $W_{\rm v, ion} < \omega_{\rm{v}, local}$, we assign a secondary MS association (labeled as ``UMS") to the absorber. We group the primary and secondary MS-associated absorbers as the Magellanic absorbers and proceed with further analyses in later sections. The remaining absorbers that do not pass any criterion are labeled as ``uncertain".

\section{UV Absorber Results} \label{sec:absorber_result}
\subsection{Associations of UV Absorbers}
Table \ref{tb:qso} and Figure \ref{fig:qso_assoc_diagram} show all absorbers with their identified associations: MS-associated (``MS", blue), galaxy-associated (``G", orange), secondary MS-associated (``UMS", light blue), and uncertain (``U", white). Each sight line is identified with a unique numerical ID (see Table \ref{tb:qso}) and is shown as a circle color-coded based on the associations of the absorbers detected along the sight line. Additionally, for absorbers that are associated with galaxies, we assign the galaxies with unique letter labels in Table \ref{tb:gal_info} and show the galaxies in Figure \ref{fig:qso_assoc_diagram}.

We consider a sight line with ``complete" associations if all absorbers along the sight line have the same associations. If a sight line contains absorbers with multiple associations, we consider it with ``partial" associations. For instance, we consider the sight line PHL2525 (ID 54 in Table \ref{tb:qso}) to have partial associations because it contains a group of ions (\CII, \CIV, \SiIII) at $v_{\rm ion}\sim-200~\kms$ that are associated with the MS, and another group of ions (\CII, \CIV, \SiII, \SiIII, \SiIV) at $v_{\rm ion} \sim-140~\kms$ that are associated with the galaxy WLM (labeled as I in Table \ref{tb:gal_info}). It is noteworthy that multiple sight lines in the vicinity of the LMC, SMC and the head of the MS have more than one association, suggesting intricate kinematics in gas within this region that is likely due to complex interplay among the LMC, SMC, the MS, and the MW's CGM.

In general, when an absorber is associated with satellite galaxies of M31, our algorithm is likely to associate this absorber with M31 itself because of the large virial radius and escape velocity. To determine which is the dominant association (M31 or its satellites), we further compare three distance-related metrics ($\Sigma_{G}$, $\delta V$, $b$), and find a handful of absorbers solely associated with M31's satellites rather than M31. For example, toward sight line IRAS-F00040+4325 (ID 25), the absorbers at velocities from $\sim-281~\kms$ to $\sim-125~\kms$ are associated with M31's satellite galaxy NGC205.

Sometimes we do not find dominant associations based on the three distance metrics ($\Sigma_{G}, \delta V, b$). For example, absorbers along sight lines HS0033+4300 (ID 22), HS0058+4213 (ID 23), IO-AND (ID 24), MRK352 (ID 42), RX-J0043.6+3725 (ID 72), RX-J0050.8+3536 (ID 73), and Zw535.012 (ID 92) are listed to be associated with M31 and either NGC185 or NGC205 or sometimes all three simultaneously.

The multiple associations of absorbers in the direction of M31
imply a spatial and kinematic proximity between M31 and its satellites NGC185 and NGC205. Yet, M33-associated absorbers are relatively well-separated in position and velocity space from M31. Among the three sight lines with absorbers associated with M33, 
all absorbers along 3C48.0 (ID 2) and RXS-J0155.6+3115 (ID 77) are completely associated with M33, while those along
MRK352 (ID 42) are partially associated. Specifically, along MRK352, absorbers at more negative velocities from $\sim-350~\kms$ to $\sim-225~\kms$ are kinematically closer to M31 but spatially closer to M33; these absorbers are considered to be associated with M31 because the kinematic proximity over-rules the spatial proximity for $\Sigma_G$ evaluation. However, absorbers at less negative velocities from $\sim-267~\kms$ to $\sim-155~\kms$ are classified to be solely associated to M33 because they exhibit lower $b$, $\delta V$, and $\Sigma_{G}$ values to M33 compared with those with respect to M31.

We treat the LMC and SMC as galaxies separate from the MS despite their known relations to examine which part of the Magellanic system the absorbers are most likely to be associated with. We find ten sight lines containing absorbers associated with either the LMC or SMC. However, only one of the ten sight lines (HE0226-4110, ID 15) has complete associations -- all the absorbers detected along this sight line are classified as associated with the SMC. For the rest of the sight lines, the spatial proximity between the LMC, SMC, and the MS causes our algorithm to classify them with partial associations, i.e., some sight lines have some absorbers associated with the LMC or SMC and others with the MS depending on the absorbers' velocities. For example, HE0435-5254 (ID 19) has absorbers with two associations: absorbers at $v_{\rm ion} \sim 130~\kms$ are associated with the MS, while those with $v_{\rm ion} >200~\kms$ are associated with the LMC. 

Apart from sight lines in the vicinity of M31 and the LMC/SMC, we also find some sight lines near nearby galaxies with absorbers with multiple/partial associations. For example, the absorbers at velocities $\sim100-180~\kms$ along HE0056-3622 (ID 13) 
are associated with NGC300, but those at negative velocities from $\sim-220~\kms$ to $-50~\kms$ are classified as uncertain (U) or with secondary MS association (UMS). 

Our algorithm assigns a significant number of absorbers to be uncertain (not associated with the MS or any galaxies). These absorbers are located closer to nearby galaxies than to the MS in terms of physical and Wasserstein distances, yet positioned outside the halos of the galaxies. Even though some of these absorbers may have been classified as associated with known structures such as the MS (e.g., \citetalias{fox20}) or the M31 (\citetalias{lehner20_amiga}) in previous studies, our algorithm provides the most conservative estimates by systematically considering a range of distance metrics (see Figure \ref{fig:flowchart}).

Many sight lines with absorbers classified as uncertain are located in the vicinity of M31 (at $90^{\degrees}\leq {\rm GLON} \leq180^{\degrees}$ and $-50^{\degrees}\leq {\rm GLAT}\leq-30^{\degrees}$), as shown in Figure \ref{fig:qso_assoc_diagram}. The presence of abundant satellites near M31 increases the probabilities of some absorbers to have lower $\Sigma_{\rm{G}}$ compared to $\Sigma_{\rm{MS}}$, causing these absorbers to be not associated with the MS (Criterion A in Figure \ref{fig:flowchart}). However, these absorbers are not considered to be associated with M31 or any of its satellites either, because the absorbers' velocities with respect to the galaxies are larger than at least one of the galaxies' escape velocities or the absorbers' impact parameters are larger than at least one of the galaxies' virial radii. As a result, these absorbers are classified as uncertain. 

Lastly, we observe that absorbers are more likely to be labeled as uncertain if they are located near complex kinematic regions. Our model relies on the average statistics of the \HI\ emission within the MS. Naturally, regions with unique kinematic or spatial complexities may not be well-represented by our model. For instance, the region near the South Galactic Pole ($270\degrees < {\rm GLON} < 360\degrees, -90\degrees < {\rm GLAT} <-70\degrees$), is unusually complex and shows the co-existence of gas structures with both positive and negative \HI\ velocities. If the \wlocalv\ were measured in this region, the velocity distribution would be bimodal. Since a bimodal distribution is atypical for most local regions in the MS, the \wlocalv\ measured from this region may not accurately represent the velocity distribution.

\begin{figure*}[t]
\includegraphics[width=\linewidth]{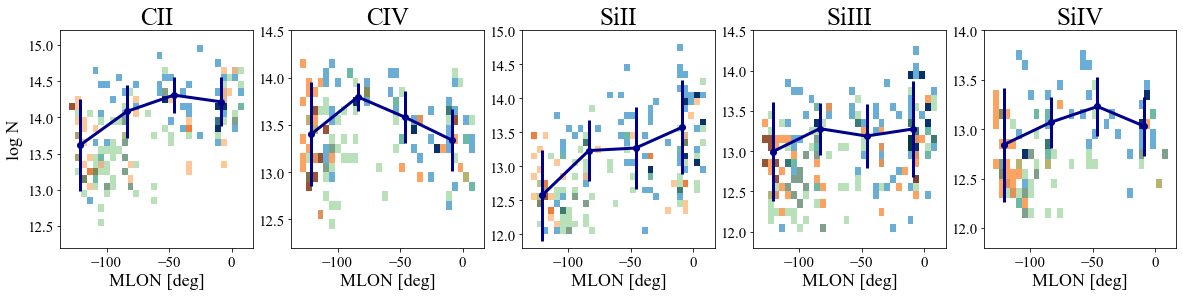}
\includegraphics[width=\linewidth]{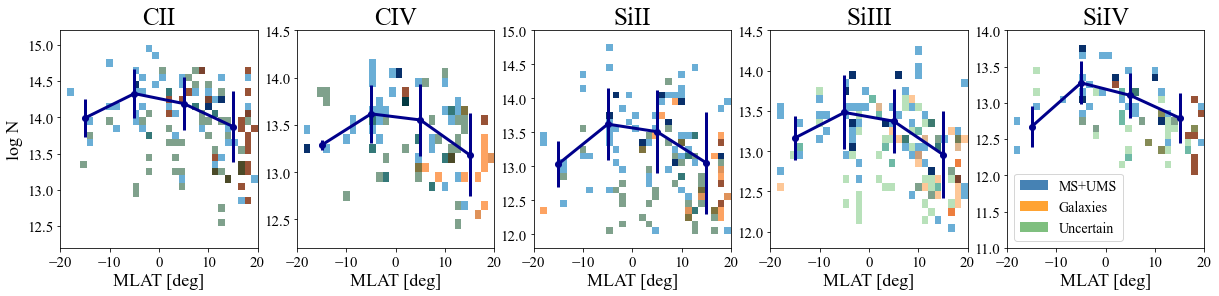}
\caption{The mean trends and 2D histograms of ion column densities as a function of Magellanic longitudes (MLON; top) and Magellanic latitudes (MLAT; bottom). The blue pixels represent averaged column densities of the MS or UMS absorbers, the orange for galaxy-associated absorbers, and the green for uncertain absorbers. The dark blue solid lines show the column densities of the MS and UMS absorbers averaged over 4.6$\degrees$ in MLON (top) and 1.3$\degrees$ in MLAT (bottom). The vertical bar indicates the standard deviation of the column densities of the MS-UMS absorbers within a given bin. The column densities of MS and UMS absorbers are roughly 0.5 dex higher than the column densities of galaxy-associated or uncertain absorbers.}
\label{fig:mean_ncol}
\end{figure*}

\subsection{Comparison to Previous Absorber Associations} \label{sec:absorber_result_compare}
In this section, we highlight the disparities between our absorber associations and the associations identified from previous studies for the same set of sight lines. 

\begin{figure*}
    \centering
    \includegraphics[width=\linewidth]{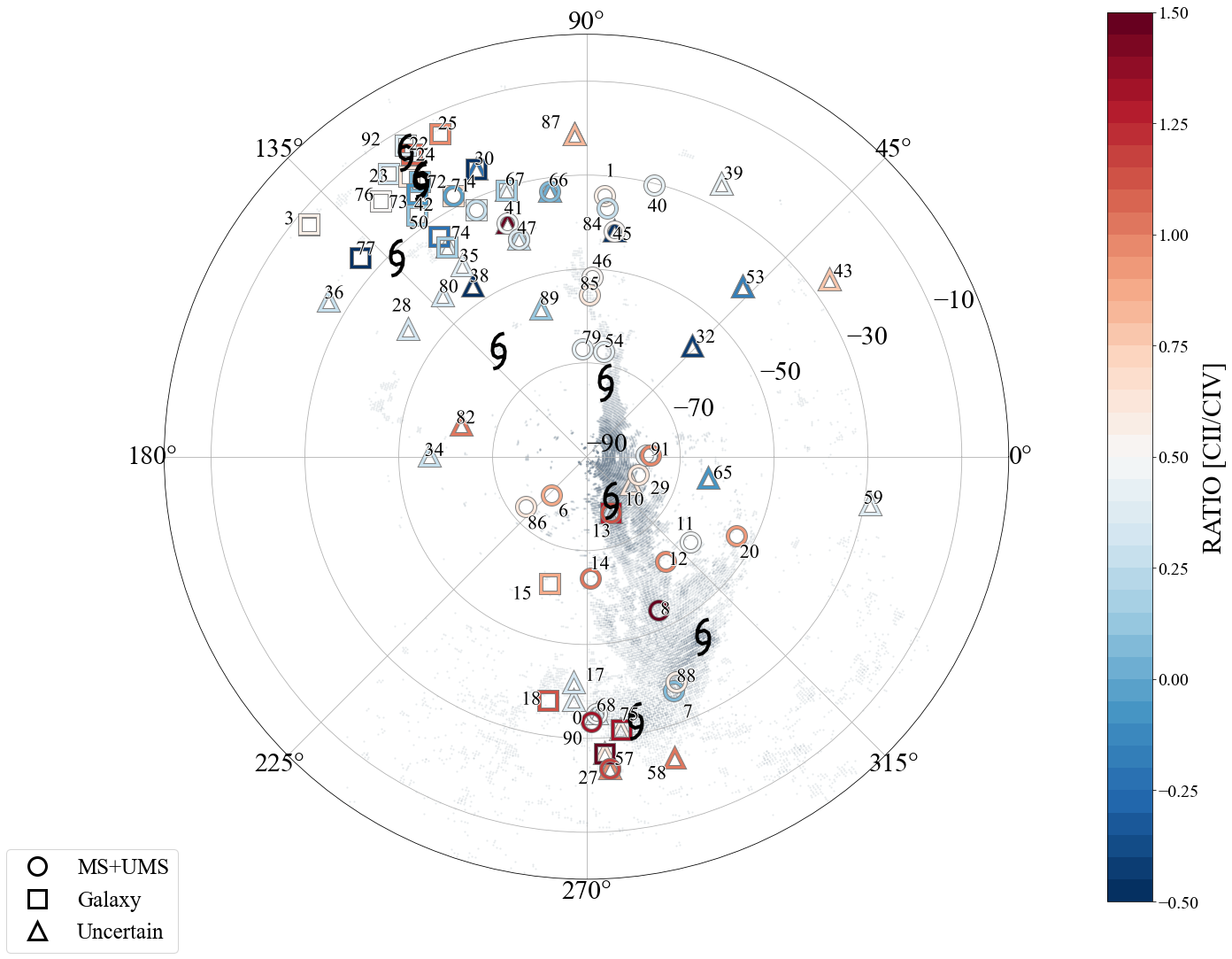}
    \caption{The spatial distribution of low-to-high ion (\CII\ to \CIV) ratios. The shape of the marker denotes the association: circles for MS-associated and UMS (secondary MS-associated) absorbers, squares for galaxy-associated absorbers, and triangles for absorbers with uncertain associations. The \CII/\CIV\ ratios of the MS and UMS absorbers generally decrease from the head region of the MS to the tail, suggesting that gas in the MS tail region is more ionized.}
    \label{fig:regional_ion}
\end{figure*}

\citetalias{fox20} associated an absorber with the MS if it is detected along a sight line within 30$\degrees$ of the 21 cm emission from the MS (defined by \citealt{morras_hi_fox14}), and the velocity of the absorber is consistent with the general kinematics of the \HI. Their sample includes 21 sight lines with MS-associated absorbers. 
Among these absorbers, we find absorbers along 11 sight lines from \citetalias{fox20}'s MS sample are either completely or partially associated to the MS. In the remaining ten sight lines, half of them are with absorbers that have uncertain associations, including MRK1044 (ID 34), MRK1513 (ID 39), PG0026+129 (ID 48), SDSSJ015530.02-085704.0 (ID 82), and UGC12163 (ID 87). The other half are with absorbers associated with nearby galaxies, including HE0226-4110 (ID 15; SMC association), IO-AND (ID 24; M31 and NGC205), LBQS0107-0235 (ID 31; IC 1613), PG0044+030 (ID 49; IC1613), and PHL2525 (ID 54; WLM). 

\citetalias{Krishnarao22} considered sight lines that are located within 45 degrees from the LMC, which corresponds to an impact parameter of 35 kpc from the galaxy. \citetalias{Krishnarao22} eliminated components linked to the MW or known intermediate-velocity clouds or high-velocity clouds by comparing absorber velocities with those of nearby known gaseous structures. In our analysis, we only consider absorber components identified as attributed to the Magellanic system in \citetalias{Krishnarao22}. Notably, all 26 sight lines from \citetalias{Krishnarao22} are either completely or partially associated with the MS, LMC, or SMC. For instance, HE0439-5254 (ID 19) absorbers with velocities below $200 \kms$ are associated with the MS, while the absorbers with velocities above $200 \kms$ are associated with LMC. While most absorbers demonstrate such partial associations with the MS and MCs, five sight lines including HE0246-4101 (ID 16), HE2305-5315 (ID 20), HE2336-5540 (ID 21), IRAS-F21325-6237 (ID 26), UKS0242-724 (ID 88) demonstrate complete association with the MS (including UMS). Although a handful of absorbers from \citetalias{Krishnarao22} are classified uncertain, all of the corresponding sight lines are partially associated with either the MS or MCs at different velocity ranges. 

\citetalias{lehner20_amiga} analyzed 43 sight lines to investigate the CGM of the Andromeda Galaxy (M31) within an impact parameter range of 25 to 569 kpc. Our algorithm shows that about half of their absorbers are associated with either M31 or its satellites, while the other half are classified differently because of their relative closeness to the MS. 
While \citetalias{lehner20_amiga} considered absorbers within $\lesssim1.9$ virial radius to be also associated with the M31, our algorithm uses a stricter criterion of one virial radius and also considers the influence of M31's satellite galaxies.

Apart from M31-associated absorbers, \citetalias{lehner20_amiga} also identified absorbers that are likely to be contaminated by the MS by examining whether the absorbers's positions and velocities are within the extent of the MS's \HI\ emission as studied by \cite{Nidever2010}. In general, these MS-contaminated absorbers are also classified as MS-associated in our algorithm, except for those along seven sight lines. We find that absorbers along sight lines
HS0033+4300 (ID 22), IO-AND (ID 24), IRAS-F00040+4325 (ID 25), KAZ238 (ID 30), RBS2055 (ID 67), and Zw535.012 (ID 92) are associated with M31 rather than the MS. On the other hand, while \citetalias{lehner20_amiga} flagged all absorbers with velocities near $\sim-370~\kms$ along RX-J0028.1+3103 as not contaminated by the MS, our algorithm identifies them as MS-associated.

Our algorithm also classifies all absorbers along 13 sight lines from \citetalias{lehner20_amiga}'s sample as uncertain, including IRAS01477+1254 (ID 28), MRK1014 (ID 33), MRK1148 (ID 35), MRK1179 (ID 36), MRK1502 (ID 38), MRK595 (ID 44), PG0026+129 (ID 48), PHL1226 (ID 52), SDSSJ011623.06+142940.6 (ID 80), SDSSJ014143.20+134032.0 (ID 81), SDSSJ015952.95+134554.3 (ID 83), UGC12163 (ID 87), and UM228 (ID 89). Additionally, our algorithm identifies absorbers along nine sight lines -- 3C454.3 (ID 1), MRK1501 (ID 37), MRK304 (ID 40), MRK335 (ID 41), NGC7469 (ID 45), PG0003+158 (ID 47), PG2349-014 (ID 51), RX-J0023.5+1547 (ID 70), SDSSJ225738.20+134045.4 (ID 84) -- from \citetalias{lehner20_amiga} to be primarily associated with the MS. 

Some sight lines have partial associations, with some absorbers associated with the M31 halo and others associated with the MS. For example, absorbers from sight line RBS2005 (ID 66) at $v_{\rm LSR}\sim-340\kms$ are classified as MS-associated, while the rest at different velocities are labeled as uncertain, indicating their proximity to the nearby galaxies. Similarly, RX-J0028.1+3103 (ID 71) has both MS and M31-associated absorbers: absorbers with velocities below $v_{\rm LSR}\sim-350~\kms$ are classified as MS-associated, while those with absorption velocities above $-350~\kms$ are associated with M31. For absorbers from sight line PG0044+030 (ID 49) in \citetalias{lehner20_amiga}'s sample, our algorithm classifies them as associated with IC1613 rather than M31.

\begin{figure*}[t]
    \centering
    \includegraphics[width=\linewidth]{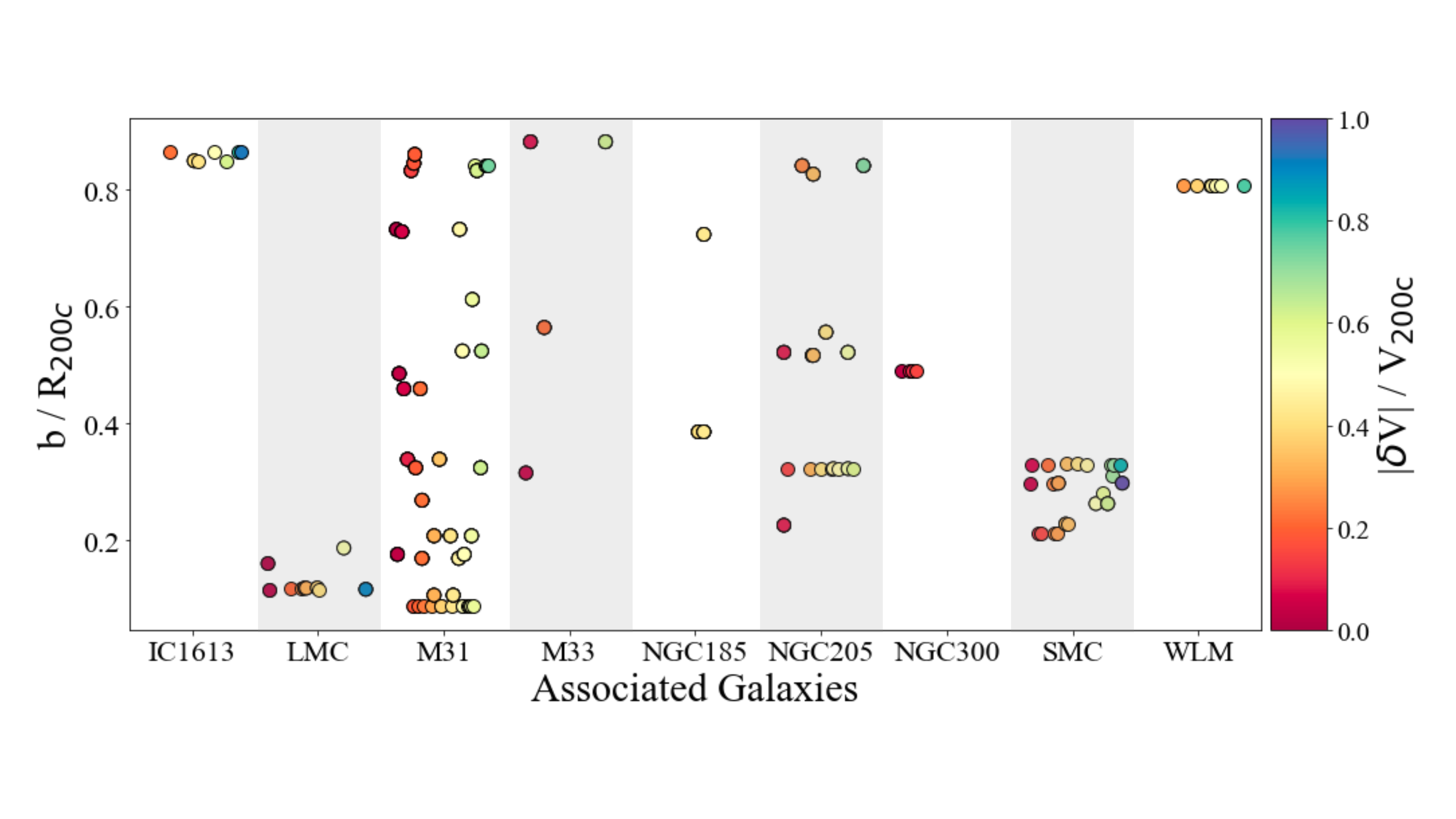}
    \caption{Impact parameter distributions of absorbers that our algorithm classifies to be associated with various galaxies. The y axis shows the impact parameters of absorbers normalized to the corresponding galaxies' virial radii. Each absorber-galaxy pair is represented by a circle, with the color indicating the velocity offset between an absorber and its associated galaxy, normalized to the galaxy's escape velocity. Note that in this Figure we use the virial radii ($R_{\rm 200c}$) and escape velocities ($V_{\rm 200c}$) defined with respect to the critical density ($\rho_{\rm 200c}$), and thus the normalized impact parameters and velocity offsets may be different from what are used for the same absorbers from the literature. 
    }
    \label{fig:impact}
\end{figure*}

\subsection{Physical Conditions of Magellanic Stream}\label{sec:ms}
In this section, we investigate the physical conditions of the MS using the MS-associated absorbers classified by our algorithm. We consider both the primary and secondary MS-associated absorbers as MS absorbers. We consider the MS absorbers in the Magellanic coordinate system, which sets the equator along the spine of the MS parallel to a great circle in the Galactic coordinate system \citep{Nidever08}. The Magellanic longitude, MLON, describes the location of an absorber along the MS from the head (near the LMC and SMC) to the tail. And the Magellanic latitude, MLAT, describes the distance of an absorber from the MS perpendicular to the long axis of the MS. 
The MS is mostly confined at MLON $\leq 0$ and $-20 \leq$ MLAT $\leq 20$, and MLON decreases along the MS (toward its tail).

Figure \ref{fig:mean_ncol} shows the column densities of absorbers as a function of MLON (top row) and MLAT (bottom row). In the background we show 2D histograms representing the column densities of absorbers for the selected ion, with different colors denoting their associations: blue for MS absorbers, orange for galaxy-associated absorbers, and green for uncertain absorbers. The blue solid line represents the binned average trend for the MS absorbers, with vertical bars indicating the standard deviations of column densities within the corresponding bins. We apply a minimum two counts threshold per each bin. Notably, the mean column densities of the MS absorbers are $\sim0.5$ dex higher than those of non-MS absorbers across the entire length and width of the MS.

The top panels of Figure \ref{fig:mean_ncol} also show that the column densities of low ions like \CII\ and \SiII\ decrease toward the Magellanic tail (more negative MLON values), while the column densities of \CIV\ tend to increase toward the tail. These trends indicate that gas in the tail region is likely to be more highly ionized. While the binned averages for the MS absorber may appear to decrease at MLON $<$ -120, this trend is influenced by only a few MS absorbers, as indicated by a high standard deviation. The mean column densities of the MS-associated absorbers remain relatively flat for \SiIII\ and \SiIV.

Figure \ref{fig:regional_ion} shows the spatial trends of the low-to-high ion ratios, \CII/\CIV, which we calculate only for absorbers adopted from the same data reference and with centroid velocity differences less than the COS resolution (25~$\kms$). We show MS absorbers as circles, galaxy-associated absorbers as squares, and those with uncertain associations as triangles. Near the head of the MS, we find that the MS absorbers exhibit higher ion ratios, which is consistent with the trend seen in Figure \ref{fig:mean_ncol} that the column densities of \CII\ are higher toward the head. The ion ratios of MS absorbers decrease toward the tail with more negative MLON, consistent with \cite{BH2019}. This trend is driven by higher \CIV\ column densities at the tail of the MS as evident in Figure \ref{fig:mean_ncol}. We observe a similar trend with low-to-intermediate ion ratios (such as \SiII / \SiIV), but with more scatter.

We calculate centroid velocity offsets for different pairs of ions, including the low-high ion pairs of \CII--\CIV\ and \SiII--\CIV, and the low-intermediate ion pairs of \CII--\SiIV\ and \SiII--\SiIV\ to understand whether these ions trace gas in the MS with similar kinematics. 
The centroid velocity offsets are computed for pairs of MS absorbers from the same data reference, and we only consider absorbers with velocity offsets within $50~\kms$ from each other to avoid outliers. 
We find that the distributions of velocity offsets for all ion pairs are centered around zero, and the widths of the distributions are consistent among ion species within the same low-high or low-intermediate groups. The velocity offset distributions between low-to-intermediate ions are narrower than those of low-to-high ions, consistent with the findings by \citeauthor{fox20} (\citeyear{fox20}; see the left panel in their figure 6).

\startlongtable
\begin{deluxetable*}{cccccc}
\tabletypesize{\footnotesize}
\tablecaption{Associated QSO-Galaxy Pairs}
\tablehead{\colhead{Galaxy Name} & \colhead{QID}  & \colhead{QSO Name} & \colhead{Ions} & \colhead{$b$}  & \colhead{$\Bar{|\delta V}|$} \\[-0.25cm]
\colhead{} & \colhead{} & \colhead{} & \colhead{} & \colhead{[kpc]} &  \colhead{[km/s]} 
\\[-0.1cm]
\colhead{(1)} & \colhead{(2)} & \colhead{(3)}& \colhead{(4)} & \colhead{(5)} & \colhead{(6)}} 

\startdata 
IC1613 & 31 & LBQS0107-0235  & \SiIII\ &  61.9 & 11.7 \\
& & &  \CII\ &  & 26.4 \\
& & & \CII, \SiIII\ &  & 45.7 \\
& 49 & PG0044+030  & \CII, \SiII, \SiIII, \SiIV\ & 60.7 & 21.8 \\ \hline
LMC & 18 & HE0435-5304 & \CIV\ & 15.3 & 18.8 \\
& & & \CII, \SiII, \SiIII\ &  & 28.6 \\
& 19 & HE0439-5254 & \CIV, \SiII\ & 15.4 & 25.8 \\
& & & \CIV, \SiIV\ &  & 37.1 \\
& 56 & PKS0355-483  & \SiII, \SiIII\ &  20.7 & 0.2 \\
& 64 & RBS1992 & \SiII, \SiIII\ & 24.2 &  50.2 \\
& 69 & RBS567  & \SiII, \SiIII\ & 15.1 &  1.5 \\
& & & \SiII\ & & 38.4 \\
& & & \CII, \SiII, \SiIII\ & & 86.8 \\ 
\hline
M31 & 3 & 3C66A & \CII, \CIV, \SiII, \SiIII, \SiIV\ & 248.0 & 38.4\\
& & & \SiIII\ & & 127.4 \\ 
& & & \CIV\ & & 154.9 \\ 
& & & \CII, \SiII\ & & 159.9\\
& 4 & 4C25.01 & \CII, \CIV, \SiII, \SiIII, \SiIV\ &  213.7 & 14.1 \\
& 9 & FBS0150+396 & \CIV, \SiII, \SiIII, \SiIV\ & 179.7 & 120.9 \\
& 22 & HS0033+4300 &  \CIV, \SiII, \SiIII, \SiIV\ & 31.3 & 68.4 \\
& & &  \CII, \CIV, \SiII, \SiIII, \SiIV\ & & 94.1 \\
& 23 & HS0058+4213  & \CII, \CIV, \SiII, \SiIII, \SiIV\ & 49.8 & 48.4\\
& & & \CII, \CIV, \SiII, \SiIII, \SiIV\ & & 100.9 \\ 
& 24 & IO-AND  & \CII, \CIV\ & 25.6 & 38.4\\ 
& & & \SiII\ & & 43.2 \\
& & & \SiIII\ & & 50.4 \\
& & & \SiIV\ & & 63.9 \\
& & & \SiII, \SiIII\ & & 80.8 \\
& & & \SiIV\ & & 91.8 \\
& & & \CII, \CIV\ & & 106.7 \\
& & & \SiII\ & & 114.0 \\
& & & \CII, \CIV\ & & 118.36 \\
& & & \SiIV\ & & 123.6 \\
& 25 & IRAS-F00040+4325  & \CII, \CIV, \SiII, \SiIII, \SiIV\ & 95.2 & 41.1 \\ 
& & & \CII, \CIV, \SiII, \SiIII, \SiIV\ & & 136.6 \\ 
& 30 & KAZ238  & \CII, \CIV, \SiII, \SiIII\ & 153.8 & 105.9 \\
& & & \CII, \CIV, \SiII, \SiIII\ & & 139.1 \\
& 42 & MRK352  & \CII, \CIV, \SiIV\ & 134.9 & 15.1 \\
& & & \CII, \CIV, \SiII, \SiIII, \SiIV\  & & 47.4 \\
& & & \CII, \CIV, \SiII, \SiIII, \SiIV\ & & 102.4 \\
& 50 & PG0052+251  & \CII, \CIV, \SiII, \SiIII\ & 214.9 & 4.9 \\
& & & \CII, \CIV, \SiII, \SiIII, \SiIV\ & & 103.4 \\
& 67 & RBS2055 & \CII, \CIV, \SiII, \SiIII, \SiIV\ & 244.3 & 30.9 \\
& & & \CII, \CIV, \SiII, \SiIII, \SiIV\ & & 134.1\\
& 71 & RX-J0028.1+3103  & \CII, \CIV, \SiII, \SiIII, \SiIV\ & 142.5 & 8.4 \\
& 72 & RX-J0043.6+3725 & \CII, \CIV, \SiII, \SiIII, \SiIV\ & 51.8 & 6.6 \\
& & & \CII, \CIV, \SiII, \SiIII, \SiIV\ & & 108.4 \\
& 73 & RX-J0050.8+3536 & \CII, \CIV, \SiII, \SiIII, \SiIV\ & 78.9 &  48.4 \\
& 74 & RX-J0053.7+2232 & \CII, \CIV, \SiII, \SiIII, \SiIV\ & 252.5 & 40.9 \\
& 76 & RXS-J0118.8+3836 & \CII, \CIV, \SiII, \SiIII, \SiIV\ & 99.5 & 20.9 \\
& & & \CII, \CIV, \SiII, \SiIII, \SiIV\ & & 75.9 \\
& 92 & Zw535.012 & \CII, \CIV, \SiII, \SiIII, \SiIV\ & 61.1 & 68.4 \\ 
& & & \CII, \CIV, \SiII, \SiIII, \SiIV\ & & 91.6 \\
& & & \CII, \CIV, \SiII, \SiIII, \SiIV\ & & 118.4 \\ \hline
M33 & 2 & 3C48.0 & \CIV, \SiII, \SiIII\ & 42.7 & 4.0 \\
& 42 & MRK352 & \CII, \CIV, \SiII, \SiIII, \SiIV\ & 119.2 & 7.5 \\
& & & \CII, \CIV, \SiII, \SiIII, \SiIV\ & & 62.5 \\
& 77 & RXS-J0155.6+3115  & \CII, \CIV, \SiII, \SiIII, \SiIV\  & 76.3 & 21.5 \\ \hline
NGC185 & 22 & HS0033+4300 & \CIV, \SiII, \SiIII, \SiIV\ & 58.5 & 25.8 \\ 
& 92 & Zw535.012 & \CII, \CIV, \SiII, \SiIII, \SiIV\ & 31.2 &  24.2 \\
& & & \CII, \CIV, \SiII, \SiIII, \SiIV\ & & 25.8 \\ 
\hline
NGC205 & 22 & HS0033+4300 & \CIV, \SiII, \SiIII, \SiIV\ & 24.5 & 6.8 \\
& 23 & HS0058+4213 & \CII, \CIV, \SiII, \SiIII, \SiIV\ & 55.9  & 25.7 \\
& & & \CII, \CIV, \SiII, \SiIII, \SiIV\ & & 26.8 \\ 
& 24 & IO-AND  & \SiIV\ & 34.9 & 11.3 \\
& & & \SiIII\ & & 24.8 \\
& & & \SiII\ & & 32.0 \\
& & & \SiIII\ & & 35.8 \\
& & & \CII, \CIV\ & & 36.8 \\
& & & \SiII\ & & 38.8 \\
& & & \CII, \CIV\ & & 43.2 \\
& & & \SiIV\ & & 48.4 \\
& 25 & IRAS-F00040+4325 & \CII, \CIV, \SiII, \SiIII, \SiIV\ & 91.0  & 19.8 \\
& & & \CII, \CIV, \SiII, \SiIII, \SiIV\ & & 58.2 \\ 
& 72 & RX-J0043.6+3725 & \CII, \CIV, \SiII, \SiIII, \SiIV\ & 60.2 & 33.2 \\
& 73 & RX-J0050.8+3536 & \CII, \CIV, \SiII, \SiIII, \SiIV\ & 89.4 & 26.8 \\
& 92 & Zw535.012 & \CII, \CIV, \SiII, \SiIII, \SiIV\ & 56.5 &  6.8 \\ 
& & & \CII, \CIV, \SiII, \SiIII, \SiIV\ & & 43.2 \\ \hline
NGC300 & 13 & HE0056-3622 & \CII, \SiIII\ & 63.8 & 1.0 \\
& & & \CIV\ &  & 3.0 \\
& & & \SiII, & & 6.2 \\  \hline
SMC & 14 & HE0153-4520 & \SiIII\ & 29.7 & 45.3 \\
& 15 & HE0226-4110 & \CII\ & 35.0 & 5.4\\
& & & \CIV\ & & 16.8 \\
& & & \SiII\ & & 26.6 \\
& & & \SiIII\ & & 32.1 \\
& & & \CII\ & & 36.3 \\
& & & \SiIII\ & & 53.1 \\
& & & \CII\ & & 55.2 \\
& & & \CII\ & & 65.4 \\
& 18 & HE0435-5304 & \CII\ & 31.5 & 3.9 \\
& & & \SiII\ & & 19.1 \\
& 56 & PKS0355-483  & \CIV\ & 33.0 &  54.0 \\
& 57 & PKS0552-640 & \CIV, \SiIV\ & 27.8 & 39.7 \\
& & & \CII, \SiII, \SiIII\ & & 48.8 \\
& 68 & RBS563  & \CII, \SiII, \SiIII\ & 24.1 & 26.6 \\
& 69 & RBS567  & \CII, \SiII, \SiIII\ & 31.5 &  22.4 \\
& & & \SiII, \SiIII\ & & 85.6 \\
& 75 & RX-J0503.1-6634 & \CIV\ & 22.4 & 8.4 \\
& & & \SiII\ & & 11.4 \\
& & & \CII\ & & 19.4 \\
& & & \SiIII\ & & 21.9 \\
\hline
WLM & 54 & PHL2525 & \CII, \CIV, \SiIII, \SiIV\ & 46.6 & 21.2 \\
& & & \CIV, \SiII & & 16.4 \\
\hline
\enddata
\smallskip

\tablenotetext{}{Note: Col (1): Galaxy names. Col (2): Unique IDs assigned to QSO sight lines as listed in Table \ref{tb:qso}.
Col (3): QSO names. 
Col (4): Ion absorbers detected along the QSO sight lines.
Col (5): Impact parameters in unit of kpc.
Col (6): Absolution velocity difference in LSR between an ion absorber and the systemic velocity of its associated galaxy.  
If multiple absorbers of the same ions are detected along the same sight lines and they are with centroid velocities within $1~\kms$, the velocity difference reflected the mean of these absorbers. 
}
\label{tb:gal_assoc}
\end{deluxetable*}

\section{Discussion} \label{sec:discussion}
\subsection{Association with Galaxies}\label{sec:galaxies}
In this paper we have gone beyond simple virial radius and velocity offsets for associating absorbers to galaxies. Our approach is primarily guided by relative comparisons, and all details are found in Sections \ref{sec:absorber_gal} and \ref{sec:determine_assoc}.
Figure \ref{fig:impact} presents the distribution of absorbers' impact parameters $b$ normalized by the corresponding galaxy's virial radius ($b / R_{200c}$) for individual associated galaxies listed in Table \ref{tb:gal_info}. Each circle represents an ion absorber with its color showing the 
velocity offset between the absorber and the galaxy's systemic velocity, normalized by the galaxy's escape velocity ($V_{200c}$). Approximately 60$\%$ of the galaxy-associated absorbers are found in the CGM of M31. This is likely due to M31 being the most massive nearby galaxy, and also the fact that at least half of the absorbers in our sample are adopted from \citetalias{lehner20_amiga}, which primarily focused on M31. In Table \ref{tb:gal_assoc}, we list the impact parameters $b$ and velocity offsets $\delta V$ for all absorber-galaxy pairs shown in Figure \ref{fig:impact}.
In the subsequent sections, we discuss individual galaxies and their associated absorbers.

\subsubsection{M31 and its Satellites}
\label{sec:m31_discuss}
The Andromeda galaxy (M31) is the closest massive galaxy to our Milky Way (located approximately 700 kpc apart) and is home to a rich system of satellite galaxies. Tidal features observed in M31's vicinity suggest that it has undergone multiple significant interactions with its satellites \citep{ngc205_m31, m31interaction_mcconnachie}. Although M31 boasts a variety of dwarf companions, we exclude the majority of M31's dwarf companions due to their lack of \HI\ gas \citep{putman21}. Two dwarf elliptical companions of M31, NGC185 and NGC205, have only small amounts of gas relative to their total mass and given their proximity to M31 are identified to be potentially associated with some absorbers. 

NGC205 lies at a projected distance of only 37 arcmin ($\sim9$ kpc) from M31 and shows evidence of distortion in its outer isophotes that are presumed to be the result of tidal interaction with M31 \citep{hodge_ngc205, sato86_ngc205, choi_ngc205}. Another dwarf companion, NGC185, lies at a projected distance of 7$\degrees$ ($\sim140$ kpc) from M31 and shows less evidence for dynamical interactions with M31 \citep{held92_ngc185}, but the star formation history of NGC185 implies an earlier infall time into the M31 environment \citep{geha15_ngc185}. A recent proper-motion (PM) measurement shows the connection between NGC185 and M31 through alignment of rotation planes \citep{paw21_ngc185}.

A vast majority of galaxy-associated absorbers near M31 are classified to be associated with M31. With \citetalias{lehner20_amiga}'s deliberate selection of sight lines, systematically probing azimuthal variations near M31, the absorbers associated with M31 and its satellites demonstrate good coverage in both impact and velocity ratios compared to other galaxies. Although not all \citetalias{lehner20_amiga} absorbers are M31-associated as mentioned in Section \ref{sec:absorber_result_compare}, all M31-associated absorbers, except IO-AND (ID 24) from \citetalias{fox20}, are originated from \citetalias{lehner20_amiga}. Details of our association comparisons with \citetalias{lehner20_amiga} is found at Section \ref{sec:absorber_result_compare}.

\begin{figure*}
\centering
   \includegraphics[width=0.95\linewidth]{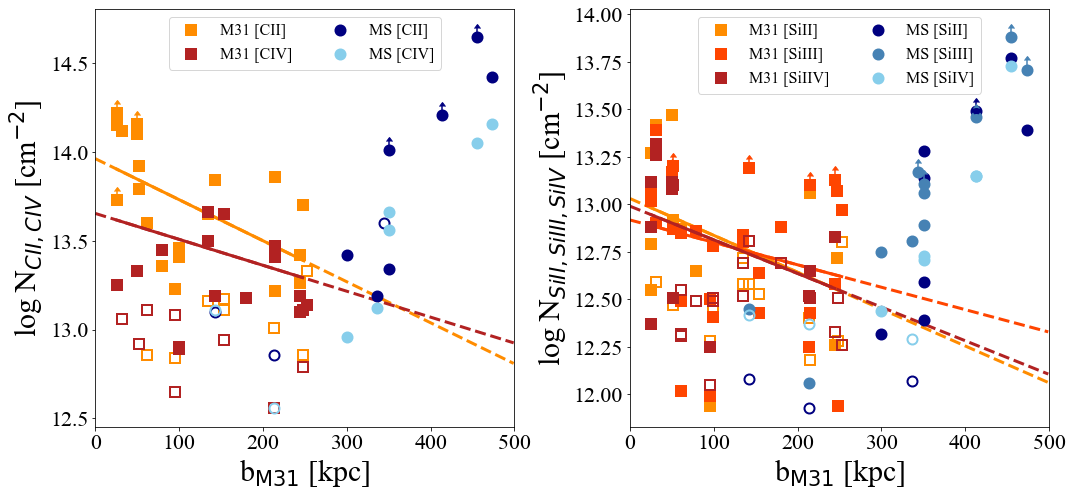}
   \caption{Ion column densities as a function of impact parameters from M31 $b_{\rm M31}$, with Carbon ions on the left, Silicon ions on the right. The red and orange squares indicate M31-associated absorbers, and the light and dark blue circles indicate MS-associated absorbers. We show lower-limit measurements as filled symbols with upward-pointing arrows, and upper-limit measurements as empty symbols. The lines denote linear fits to the detection and lower-limit data points associated with M31. The solid portion of a line represents the range covered by the actual data, and the dashed portion is an extrapolation to larger $b_{\rm M31}$. In general, we find ion column densities decline with $b_{\rm M31}$ for M31-associated absorbers. At $b_{\rm M31}\gtrsim R_{\rm 200c, M31}(\sim300$ kpc), the MS-associated absorbers are generally found at more negative $v_{\rm LSR}$ (not shown here) and the ion column densities increase with distances further away from M31. } 
   \label{fig:ncol_m31}
\end{figure*}

Figure \ref{fig:ncol_m31} displays the logarithmic column density values for individual ion absorbers as a function of impact parameters from M31. The M31-associated absorbers are represented with square markers, while the MS-associated absorbers are marked with circles. Filled markers denote detections or lower limits (with upward pointing arrows), while empty markers indicate upper limits. We show carbon ions on the left and silicon ions on the right. As shown, we find contrasting trends between the M31-associated and MS-associated absorbers for all available ions. In general, the column densities of M31-associated absorbers decrease as the impact parameter increases, indicating a decrease in gas content with radial distance, a signature of a diffuse circumgalactic medium (CGM) \citep{tumlinson_cgm}. In contrast, the column densities of MS-associated ion absorbers increase as the distance to the MS decreases and the impact parameter of M31 increases. Two sight lines, 4C25.01 (ID 4) and RX-J0028.1+3103 (ID 71) located at $b_{\rm M31} \sim 213$ and $142$ kpc respectively, are partially associated with both the MS and M31 with two groups of absorbers at different velocities. Our algorithm classifies their absorbers with velocities $\sim-400$ to $\sim-330~\kms$ be associated with the MS, while the rest of absorbers above this velocity range are classified as M31-associated.

To investigate the general trend further, we perform linear regression on the column densities and impact parameters of the M31 associated absorbers with only detection and lower limit measurements. In Figure \ref{fig:ncol_m31}, a solid line represents the actual data range, while a dashed line is an extrapolation based on the fit. The Pearson correlation coefficient $r$ values ranging from -0.55 to -0.21, suggesting negative correlations between the column densities and impact parameters for M31 associated absorbers. The correlation for \CII\ yields a p-value of 0.01 ($<0.05$), suggesting that the correlation is significant. However, the p-values for other ions are greater than 0.1, suggesting that given the current data sample and the large scatters we are unable to rule out the null hypothesis that there is no strong correlation between the ion (\CIV, \SiII, \SiIII, \SiIV) column densities and impact parameters related to M31. 
As shown, the slope of the \CII\ and \SiII\ profiles are the steepest and the slope of the \SiIII\ profile is the shallowest. We also find that the standard deviations of the column densities for the silicon ions are roughly twice as high as 
those for the carbon ions due to large scatter.

Similar to \citetalias{lehner20_amiga}, we find it challenging to discern a clear trend in ion ratios as a function of $b_{\rm M31}$ due to the large scatter. For absorbers along the same sight lines but from different data sets, we consider them to be different absorbers if their centroid velocities are offset by more than the COS resolution limit, regardless of the overlaps in their velocity ranges. For this reason, absorbers with wider velocity ranges would have higher column densities than those with narrower ranges, even though they are partially probing the same gas structures along the line of sight. 
The large scatter may also be caused by ion absorbers likely associated with M31 and its satellites. Although we do not detect much of a radial trend in most directions around M31, we do find a notable increase in the ion ratios in the direction of NGC185 from M31 and near M33, especially in low-to-high ion ratios (\CII/\CIV\ and \SiII/\SiIV).

\begin{figure*}
\centering
   \includegraphics[width=0.95\linewidth]{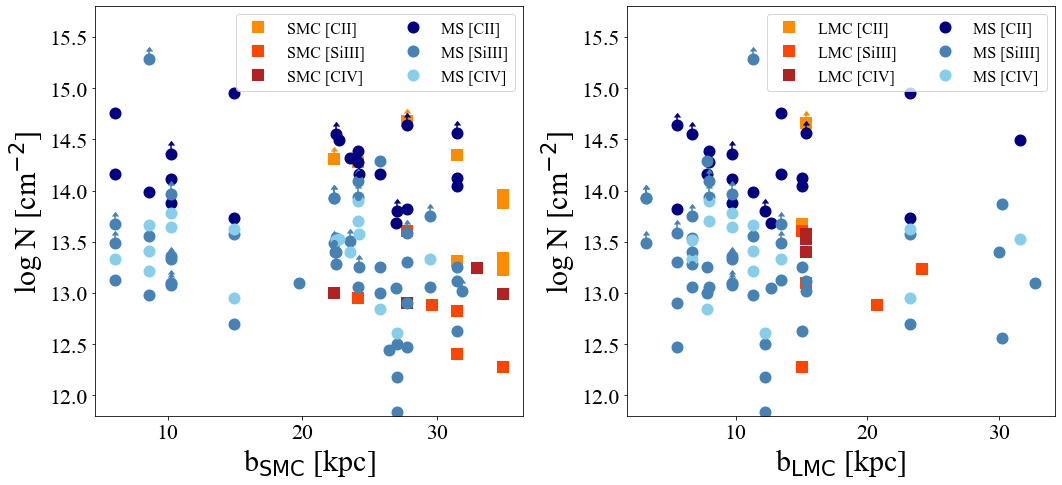}
   \caption{The column densities of ion absorbers associated with the SMC (left) and the LMC (right) as a function of the respective impact parameters. The orange and red squares indicate SMC-associated (left) or the LMC-associated (right) absorbers, and the light and dark blue circles indicate MS-associated absorbers. We show lower-limit measurements as filled symbols with upward-pointing arrows. We do not include ion absorbers with upper-limit column densities from \cite{Krishnarao22} because of the lack of velocity information in those non-detection absorbers (see \citeauthor{Krishnarao22}'s extended data table 1).
   Our algorithm shows that in the general area of the LMC, SMC, and the head of the MS, ion absorbers blend together heavily with similar column densities. There is no obvious radial trend in ion absorbers associated with the SMC or the LMC. In fact, we find that ion absorbers at small impact parameters with respect to the SMC or the LMC are kinematically more similar to the MS gas, and our algorithm classifies these absorbers as MS associated. 
   }
   \label{fig:ncol_smc}
\end{figure*}

\subsubsection{M33} \label{sec:m33}
M33 is the third most massive galaxy in the Local Group (with approximately 1/20 of the mass of M31). It is located at $\sim$190 kpc from M31 \citep{m33_corbelli} and 840 kpc from the Milky Way \citep{freedman91}, and has a velocity of $v_{\rm LSR}=-179~\kms$ \citep{m33_corbelli}. Among the absorbers that we adopt from \citetalias{lehner20_amiga}, which they assumed to be associated with M31, we find absorbers from three sight lines including 3C48.0 (ID 2), MRK352 (ID 42), RXS-J0155.6+3115 (ID 77), to be associated with M33 instead (see Figure \ref{fig:qso_assoc_diagram}). 

Figure \ref{fig:impact} shows that these M33-associated absorbers are at a range of impact parameters, but all at velocities close to the systemic velocity of M33. Compared with absorbers associated with NGC185 or NGC205, M33-associated absorbers are relatively well-separated in position-velocity space from M31. The absorbers are detected in various ions, such as \SiII, \SiIII, \SiIII, \CII, and \CIV, suggesting that M33's CGM is multiphase. All absorbers are detected over a velocity range of $-270\lesssim v_{\rm LSR}\lesssim -140~\kms$, which is consistent with the absorption of inflows detected toward stars in M33 \citep{zheng17}. The ion absorption is more extended in bluer velocities (more negative $v_{\rm LSR}$) with respect to M33; this is likely because of a sampling bias due to M33's systemic velocity ($v_{\rm LSR}=-179~\kms$) being close to the MW's high-velocity cloud absorption, in which case absorbers at less negative velocities may be omitted in \citetalias{lehner20_amiga}'s sample to avoid contamination from the MW halo gas.  The presence of a CGM out to large radii has significant implications on the likelihood of an interaction between M33 and M31 in the past.   A halo medium of M33 would likely be stripped by M31's halo medium if M33 had been within the inner halo of M31 in the past.

\subsubsection{LMC and SMC}
The Magellanic Clouds (LMC and SMC) are dwarf irregular satellites of the MW. The LMC is located at  $50\pm1$ kpc \citep{Walker_lmc_dist} from us, and the SMC is at $61\pm1$ kpc \citep{smc_dist}. The MCs are interacting with both each other and the MW, which results in a very complex gaseous environment.
Despite having different association labels, the ions from the MCs and the MS overlap significantly in the column density-impact parameter space compared with M31. Figure \ref{fig:ncol_smc} shows the logarithm of column densities of individual ion absorbers around the SMC as a function of $b_{\rm{SMC}}$ on the left and the column densities of ion absorbers around the LMC as a function of $b_{\rm{LMC}}$ on the right. We show either LMC or SMC associated absorbers with squares and mark the MS-associated absorbers with circles. As before, the lower limit measurements are indicated with upward-pointing arrows. We note that Figure \ref{fig:ncol_smc} does not include sight lines with upper-limit column density measurements that were reported by \citetalias{Krishnarao22}. This is due to the lack of velocity information for upper-limit data points in \citeauthor{Krishnarao22}'s published data (see their extended data table 1). Given that our work relies on both position and velocity information to quantify possible associations for ion absorbers, we are unable to process those absorbers with only upper-limit column density values from \cite{Krishnarao22}.

Unlike the M31 radial trend shown in Figure \ref{fig:ncol_m31}, Figure \ref{fig:ncol_smc} does not demonstrate a discernible trend between the MS-associated absorbers and those associated with the SMC (left panel) or the LMC (right panel). This is consistent with previous studies that found the MCs and the MS are part of the larger Magellanic System \citep{WW72, Mathewson74, putman2003, fox13_abundance, MS_onghia}. It is also consistent with the fact that gas emanates continuously from the SMC into the MS, and the absorbing gas is likely associated. 

\citetalias{Krishnarao22} reported a possible detection of a Magellanic corona hosted by the LMC. Their model suggests the corona of the LMC is a collisionally ionized, warm-hot gaseous halo at a virial temperature of $T\sim$10$^{5.3-5.5}$K that extends to the virial radius of the LMC ($\sim$100--130 kpc). In their analysis, \citetalias{Krishnarao22} excluded the fractions of \SiIV, \CIV, and \OVI\ column densities due to photoionization and found that the rest of the ion column densities due to collisional ionization decline as a function of impact parameters from the LMC, which they interpreted as evidence for the presence of a Magellanic corona.

Although \citetalias{Krishnarao22} considered all absorbers within 35 kpc of the LMC to be part of its corona gas, our algorithm shows that a large fraction of these \citetalias{Krishnarao22} absorbers are instead closer to either the MS or the SMC when considering the absorbers' proximity in both position and velocity space (see Section \ref{sec:absorber_result_compare} and Table \ref{tb:qso}). In fact, Figure \ref{fig:ncol_smc} shows that most absorbers within $b_{\rm SMC}\lesssim 20$ kpc of the SMC (left panel) or $b_{\rm LMC}\lesssim 15$ kpc of the LMC are kinematically closer to the MS and thus classified to be associated with the MS by our algorithm.  There is a large scatter, but absorbers in the general area of the LMC, SMC, and the MS show similar column densities and there is no obvious trend in $\log N$ \textsl{vs.} $b_{\rm SMC}$ or $b_{\rm LMC}$. Similar to what \cite{Krishnarao22} suggested, our result finds that the absorber environment in this area is complex and ion absorbers from the LMC's corona, if it exists, may not be as cleanly separated from the MS and SMC gas.

\subsubsection{NGC~300}
The Sculptor group is one of the nearest galaxy groups beyond the Local Group, with the distances of its members ranging from approximately 2 to 5 Mpc \citep{sculptor_karachentsev, jerjen98}. The proximity results in low galaxy velocities that overlap with those of the Magellanic Stream \citep{putman2003}.  This is particularly a problem at the near end of the group where NGC300 and NGC55 are found, and this has prohibited previous attempts to identify associated absorbers or \HI\ clouds. 

With our algorithm, we are able to identify one sight line (HE0056-3622, ID 13) that has absorbers associated with NGC300, which is a late type spiral with a very extended \HI\ disk \citep{westmeier11}. As shown in Figure \ref{fig:impact},  we find \CII, \CIV, \SiII\ and \SiIII\ absorbers at approximately 0.5 R$_{200c}$ and with velocities very close to the systemic velocity of the galaxy. The column densities detected are typical of those found for the CGM of star forming galaxies \citep[e.g.][]{prochaska17}, although the stellar mass of $\sim10^9~M_\odot$ places it more in the dwarf galaxy regime. The values are on average $\sim0.5-1.0$ dex higher than is typically found for dwarf galaxies at similar impact parameters \citep{zheng23}. \cite{westmeier11} note that the origin of the unusually extended gaseous disk may be due to an interaction with another galaxy (or medium) in the Sculptor group, and this could result in more gas being present in the galaxy's halo.

\subsubsection{IC1613}
IC1613 is a dwarf irregular galaxy located on the outskirts of the Local Group. Because of its isolated location, IC1613 is unlikely to be affected by ram pressure stripping from nearby massive systems like the MW or M31 \citep{Zheng_ic1613}. While \cite{Zheng_ic1613} studied 6 QSO sight lines near IC1613 at projected distances of $\sim$6-61 kpc, we only adopt two (LBQS0107-0235, PG0044+030) sight lines that were included in \citetalias{fox20}'s Magellanic sample. For these two sight lines, our algorithm identifies the absorbers near IC1613's systemic velocity ($V_{\rm{LSR}} \sim -236~\kms$; IDs 31 and 49) to be associated with the galaxy, consistent with what \cite{Zheng_ic1613} has found. We note that \citetalias{fox20} originally classified the absorbers to be associated with the MS; this is likely because they did not consider background galaxies in close spatial and velocity projection to the MS. As for column density measurements, we find that \cite{Zheng_ic1613}'s \CII\ and \SiIII\ logN values are consistent with \citetalias{fox20}'s when considering absorption near IC1613's systemic velocity, regardless of the methods being used (either AOD or VP). However,  \cite{Zheng_ic1613} also detected \SiII\ and \CIV\ absorbers of LBQS0107-0235 near IC1613, but these were not reported by \citetalias{fox20}. The reason for the missing measurements by \citetalias{fox20} is unclear because we have confirmed that these two ion absorbers appear to be well detected based on the published spectra by \cite{Zheng_ic1613} (see their Figure A12).

For the sight line PG0044+030, we also collect absorber data from \citetalias{lehner20_amiga} and compare the values with \cite{Zheng_ic1613}'s. As shown in Table \ref{tb:qso} (ID 49), two velocity groups can be found in various ions: one from $v_{\rm LSR} \sim-350$ to $\sim-250~\kms$ that our algorithm identified as U (uncertain), and the other one from $v_{\rm LSR} \sim-250~\kms$ to $\sim-170~\kms$ that we identify to be associated with IC1613. In \cite{Zheng_ic1613}, the two velocity groups can be seen in \CII\ and \SiIII, the spectra of which were fitted with Voigt profiles; in this case, the absorber from $v_{\rm LSR} \sim-250~\kms$ to $\sim-170~\kms$ is considered to be more likely to be associated with IC1613's CGM\footnote{For the velocity component from $v_{\rm LSR}\sim-350$ to $\sim-250~\kms$, \cite{Zheng_ic1613} flagged the \CII\ absorber as ``non-association" (with IC1613), which is consistent with the U (uncertain) flag in our algorithm. However, they flagged the \SiIII\ at similar velocity as associated with IC1613 while we flag it as U (uncertain); this is likely due to the different escape velocity that we adopt for IC1613's halo as compared to that of \cite{Zheng_ic1613}}. For \SiII\ and \SiIV\ where only AOD measurements are available, \cite{Zheng_ic1613}'s velocity integration ranges are slightly narrower from the two velocity groups combined, but the total column densities are similar when considering the uncertainties. 

To summarize, our algorithm shows similar membership classification as \cite{Zheng_ic1613}'s in the two QSO sight lines (LBQS0107-0235 and PG0040+030) at impact parameters of $\sim60$ kpc from the galaxy. Even though these absorbers were originally classified as associated with the MS (\citetalias{fox20}) or M31's CGM (\citetalias{lehner20_amiga}), our work demonstrates that the overall environment of an absorber -- foreground Stream and nearby galaxies in close projection -- should be thoroughly considered before assigning associations.

\subsubsection{WLM}
WLM is a gas-rich dwarf irregular (dIrr) galaxy located on the outskirts of the Local Group at a distance of 0.93 $\pm$ 0.03 Mpc \citep{McConnachie_dwarf}. Like IC1613, WLM is an isolated system whose nearest neighbor (Cetus dwarf spheroidal galaxy) is at a distance of $\approx$ 210 kpc. We find one QSO sight line, PHL2525 (ID 54 in Table \ref{tb:qso}), that is located at an impact parameter of 45.2 kpc ($\sim0.8~R_{\rm 200c}$\footnote{Note that we adopt a different virial radius definition from what \cite{Zheng_WLM} used in their work.}) from WLM. Along this sight line, there are two main absorber groups: one centers at $v_{\rm LSR} \sim-150\kms$ and another at $v_{\rm LSR} \sim-210\kms$. Our algorithm shows that the absorbers at $\sim-150\kms$ are associated with WLM, while the other group of absorbers at $v_{\rm LSR} \sim-200\kms$ are associated with the MS, which is consistent with what \cite{Zheng_WLM} found in their study of the WLM's CGM. 

\subsection{Implications on the Magellanic System} \label{sec:ms_discuss}
In the Magellanic coordinate system, all of our Magellanic-associated absorbers are located between $-118\degrees \leq$ MLON $\leq 6\degrees$ and $-28\degrees \leq$ MLAT $\leq 22\degrees$. Out of a total of 605 Magellanic absorbers in the area mentioned above, 248 absorbers are associated with the MS, 61 absorbers are assigned with the secondary association to the MS (UMS), 209 absorbers are labelled uncertain, and the rest is galaxy-associated. We find a covering fraction of 78$\%$ when we exclude uncertain absorbers from the calculation. 

Based on our association and UV absorber detection rate, we estimate the cross-section of the Magellanic gas using the same method used by \cite{fox14}. By multiplying the covering fraction and the area on the sky in which the Magellanic UV absorbers are detected, we obtain a total cross-section of $\sim 4820$ deg$^2$. In contrast, \cite{fox14} estimated the effective area of the MS by summing the area of the non-shaded grid cells in their figure 1b. Since \cite{fox14} considered the entire Magellanic system, we capture the relevant area from their figure for the comparison and multiply by their covering fraction 81$\%$ to estimate their relevant cross-section as $\sim 7047$ deg$^2$. Because of the strict association algorithm that we implement in this work and our MS extent is inferred by \HI\ emissions, the Magellanic cross-section that we derive is only $\sim70\%$ of what \cite{fox14} suggested. Assuming that the cross section of the whole Magellanic System (including the Leading Arm) is reduced by the same rate, our work suggests that the total ionized mass of the Magellanic Stream inferred by \cite{fox14} should be scaled down by $\sim30\%$. We note that our value may grow if deeper HI observations significantly extend the HI emission of the MS \citep[e.g.][]{stanimirovic08,westmeier08}.

Our results are consistent with the finding that the MS contains a substantial amount of ionized gas, and also highlight some unique aspects of the ionized gas. Figure \ref{fig:mean_ncol} suggests the mean ion column density of the MS is generally higher ($\sim$0.5 dex) than absorbers not associated by our algorithm with the MS in all directions. With the exception of \CIV, the region with the highest ion column densities coincides with the two \HI\ filamentary structures connecting the SMC and the LMC, with a gradient along the length of the MS. This gradient is also seen in the \HI\ column densities \citep{Mathewson77, mirabel_1981, putman2003, parkes_hi_mcs} and strength of $H\alpha$ emission \citep{putman03_halpha, barger_halpha} along the MS. The gradient is more pronounced in the low ions, consistent with the colder gas representing the more recently stripped material from the Magellanic Clouds and the gas being more highly ionized further along the MS. 

From Figure~\ref{fig:regional_ion}, we see direct evidence of the gas associated with the MS being more ionized toward the tail of the MS. We also examine a similar trend for other ion ratios, but the gradient is not as obvious as for \CII/\CIV. 
The increased amount of highly ionized gas toward the MS tail is likely due to the gas there being more fragmented and optically thin, which is consistent with previous findings that the \HI\ debris clouds near the tail are likely to represent the cold peaks of a largely ionized structure \citep{westmeier08, Kumari_mcs}. The H$\alpha$ emission along the MS is too bright in regions to be sourced by photoionization from the MCs, MW or the extra-galactic background \citep{putman03_halpha, barger_halpha,BH13, fox20}. The ionization of the MS is likely due to a combination of its interaction with another medium and radiation from the Galactic disk \citep{Moore94, BH07}.
Future work on the ion ratios, combined with deeper HI emission observations, should provide further insights into the ionization mechanisms of the MS.

\section{Summary}
In this work, we present a method to identify the associations of \HI\ emission elements and multiphase UV absorbers in the Magellanic Stream (MS) and its surroundings. We examine three distinct types of associations: the association of the \HI\ emission that is part of the MS, the association of an ion absorber with the \HI\ of the MS, and the association between an absorber and any nearby galaxies in close projection. To facilitate these associations, we assemble absorber data from 92 sight lines from the literature, archival and new COS observations, an \HI\ emission map of the MS, and a sample of local galaxies in close projection to the MS that contain \HI\ gas. We summarize our key findings below.

\begin{enumerate} 
    \item We construct a distance metric $\Sigma$ based on the Wasserstein distance (WD) to quantify associations. We assess the levels of \HI\ associations through score functions, which determine the membership probabilities of individual \HI\ elements. We evaluate the WD between each \HI\ element and the entire MS in both spatial and velocity spaces (\wglobalv\ and \wglobalx) and fit probability distribution functions (PDF) to the resulting \wglobalv\ and \wglobalx\ distributions. 
    
    To mitigate biases due to density-based association, 
    we determine the \HI\ association by evaluating the score based on the WDs computed between an \HI\ element and its kinematic and spatial local regions (\wlocalv\ and \wlocalx). We observe a discernible break in both score distributions at 0.9, at which we define the corresponding velocity and spatial WDs, $\sim$117.7 $\kms$ and 25 degrees, as the characteristic geometric distances of the MS. Using these values as normalization constants, we define a comprehensive distance metric $\Sigma$ as depicted in Equation \ref{eq:sigma}.

    \item By utilizing an \HI-specific distance metric $\Sigma_{\rm{HI}}$ and the derived spatial and velocity \HI\ scores, we examine the associations of individual \HI\ elements to the main body of the MS. Low velocity scores signify kinematically complex regions, such as those near the South Galactic pole or close to the MCs. Conversely, low spatial scores concur with fragmented clouds near the head region of the MS (see Figure \ref{fig:hi_score_distr}). We consider a threshold of $\Sigma_{\rm HI} = \sqrt{2}$ as the boundary condition for \HI\ associations, which corresponds to characteristic spatial and velocity WDs of the MS.
    
    \item 
    We compute the distance metrics $\Sigma_{\rm MS}$ and $\Sigma_{\rm G}$ to investigate UV absorbers' associations with either the MS or nearby galaxies in close projection, respectively. Following the procedure outlined in Figure \ref{fig:flowchart}, we determine the dominant source of association for individual absorbers. We categorize absorber associations into four groups: MS, UMS (secondary MS), galaxies, and uncertain. We compare our associations with previous studies in Section \ref{sec:absorber_result_compare} and find overall agreement, but differences in some specific absorbers. 

    \item     
    Our analysis reveals that the absorbers associated with the MS exhibit column densities approximately 0.5 dex higher than those of non-MS absorbers across all ions (\CII, \CIV, \SiII, \SiIII, \SiIV). We observe that the ion column densities tend to increase with higher MLON values (head of the MS) with the exception of \CIV, which exhibits elevated column densities at lower MLON values (tail of the MS).  The MS absorbers also demonstrate higher \CII/\CIV\ ratios and a more pronounced increasing trend in ion ratios as a function of MLON when compared with non-MS absorbers. 

    \item 
    Our algorithm also identifies a number of absorbers to be associated with nearby galaxies, including the Magellanic Clouds (LMC, SMC), M31 and its satellite galaxies (NGC185, NGC205) as well as other smaller galaxies such as M33, NGC300, IC1613, and WLM (see Figure \ref{fig:impact}). For absorbers associated with M31 within its virial radius $R_{\rm vir, M31}(\sim300)$ kpc, we find that the absorbers' column densities decrease with impact parameters, as shown in Figure \ref{fig:ncol_m31}. Beyond $R_{\rm vir}$, the absorbers' associations become less certain; our algorithm does not consider absorbers outside the virial radii to be associated with any galaxies, and for absorbers near the tail of the MS, our algorithm often classifies them to be associated with the MS. Our algorithm shows that the boundary between M31's halo and the ionized envelope of the tail of the MS may be not clearly separated, and ion absorbers in this area are likely to be blended with origins from both sources.

    \item For absorbers associated with the LMC or the SMC, we find that they are typically at impact parameters of $b_{\rm SMC}\sim20-35$ kpc or $b_{\rm LMC}\sim15-25$ kpc, as shown in Figure \ref{fig:ncol_smc}. For those absorbers within $b_{\rm LMC}\lesssim15$ kpc of the LMC, our algorithm classifies them to be dominantly associated with the MS given that these absorbers are kinematically closer to the MS than the LMC. Our analysis demonstrates that, when both spatial and kinematic similarities are considered, ion absorbers near the general Magellanic area are more likely to be associated with the MS than the SMC or the LMC. The LMC's corona gas, if it exists, is likely to be heavily blended with ambient gas at all impact parameters, similar to what \citep[e.g.][]{Krishnarao22} suggested.
    
    \item We identify new absorbers from three QSO sight lines, 3C48.0, MRK352, and RXS-J0155.6+3115, that are potentially associated with the CGM of M33, and absorbers from HE0056-3622 that are associated with the CGM of the Sculptor Group galaxy NGC300. We find overall consistency with previous studies in associations for absorbers associated with either IC1613 or WLM.
\end{enumerate}

\section*{Acknowledgements}
We are grateful for the helpful and supportive feedback from the anonymous referee. D.K. thanks helpful discussions with Hannah Bish, Hsiao-Wen Chen, Dhanesh Krishnarao, Adrian Liu, Josh Peek, Joe Suk, and Long Zhao. D.K. acknowledges Interstellar Institute's program ``II6" and the Paris-Saclay University's Institut Pascal for hosting discussions that nourished the development of the ideas behind this work. This project makes use of astropy \citep{astropy:2018}, numpy and scipy \citep{scipy}, matplotlib \citep{matplotlib}, and statsmodel \citep{statsmodel}.  This work is partially based on observations made with the NASA/ESA Hubble Space Telescope (program ID: No. 16301 and No. 15156). YZ acknowledges support from grant HST-AR-16640. Support for HST-GO-16301, HST-GO-15156, and HST-AR-16640 was provided by NASA through a grant from the Space Telescope Science Institute (STScI). STScI is operated by the Association of Universities for Research in Astronomy, Inc.,
under NASA contract NAS5-26555. This research has made use of the HSLA database, developed and maintained at STScI, Baltimore, USA.

\section*{Data availability}
This publication utilizes data from HST Science Archive (https://archive.stsci.edu). Some/all of the data presented in this article were obtained from the Mikulski Archive for Space Telescopes (MAST) at the Space Telescope Science Institute. The specific observations analyzed can be accessed via \dataset[DOI: 10.17909/6p00-pc15]{https://doi.org/10.17909/6p00-pc15}. The data that support the plots within this paper and other findings of this study are available from the corresponding author upon requests.

\appendix
\section{Analyses of Absorbers Included in our Archive Sample} \label{appendix:archive}

\begin{figure*}
\centering
\includegraphics[width=0.85\linewidth]{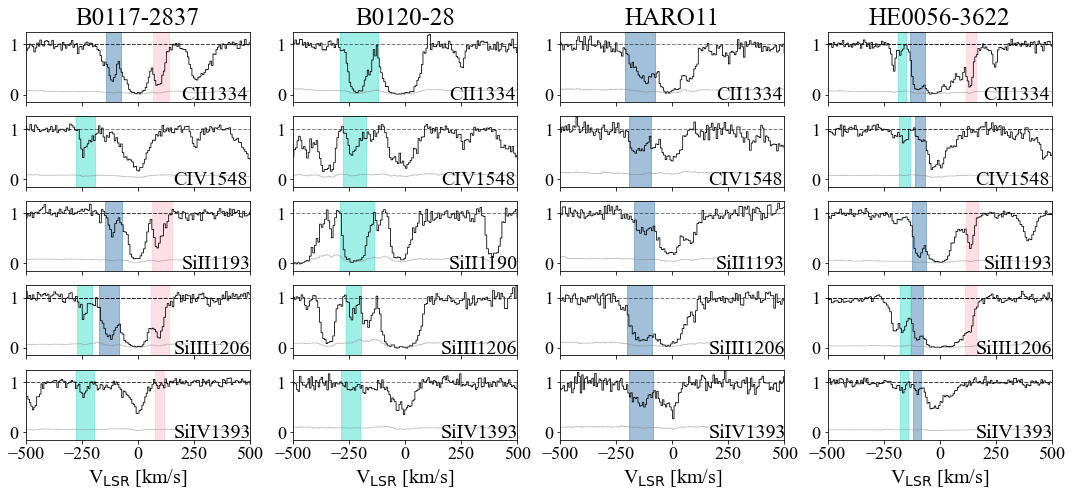}
\includegraphics[width=0.85\linewidth]{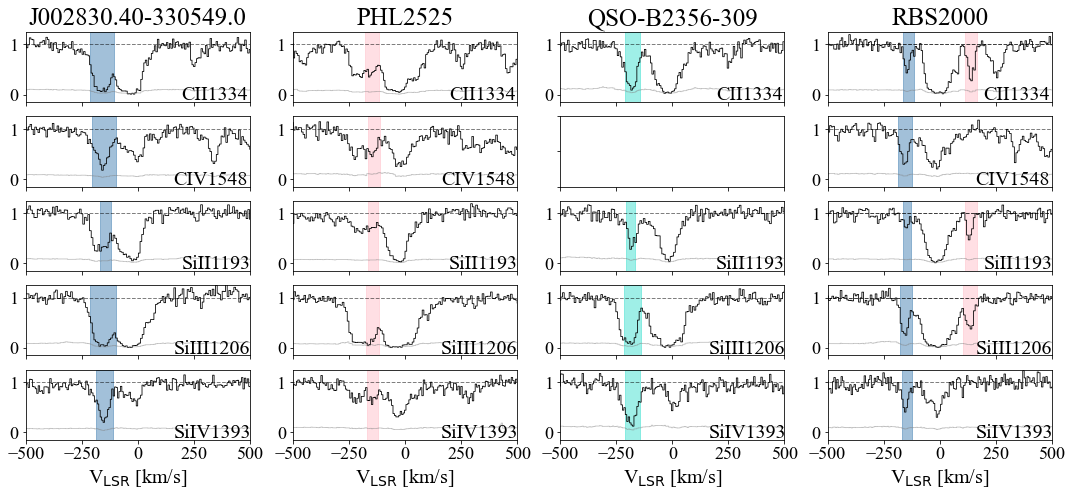}
\includegraphics[width=0.85\linewidth]{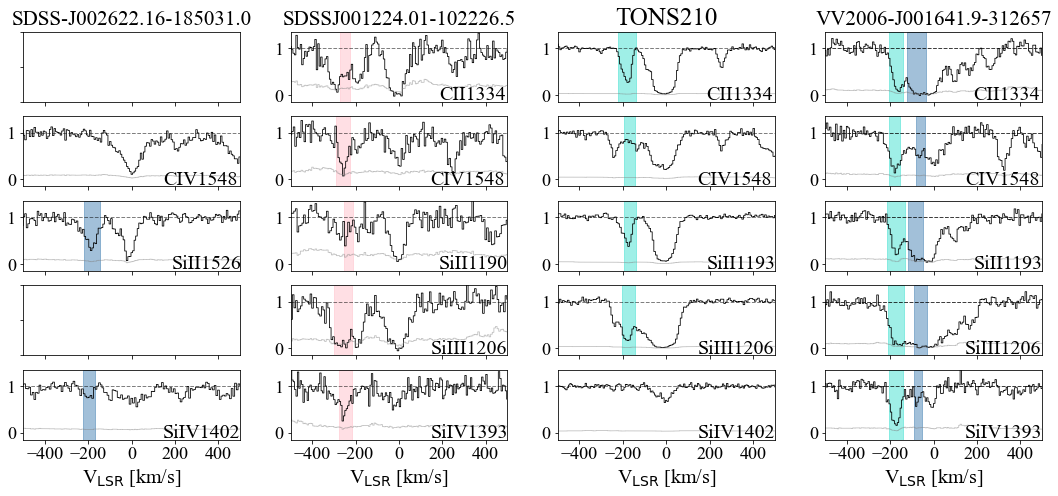}
\caption{
Normalized line spectra for QSO sight lines included in our archive sample. For absorbers for which we calculate column densities and centroid velocities based on the AOD method, we highlight their ($v_{\rm min}$, $v_{\rm max}$) velocity spans in different colors; each color represents an individual absorber identified in our algorithm. Note that along each sight line, we do not attempt to identify all absorbers with distinct velocity separations. For absorbers that have been identified in previous literature works based on the VP method (i.e., \citealt{fox20} and \citealt{Krishnarao22}), we adopt absorber measurements from the corresponding references instead and do not show these absorbers here. As mentioned in Section \ref{sec:qso_data}, we primarily consider \CII\ $\lambda$1334, \SiII\ $\lambda$1193, \SiIII\ $\lambda$1206, \SiIV\ $\lambda$1393, and \CIV\ $\lambda\lambda$1548 transition lines when available. Occasionally, we choose weaker lines (e.g., \SiII\ $\lambda$1190 or \SiIV\ $\lambda$1402) when the primary lines are either saturated or contaminated.
}
\label{fig:aod_fit}
\end{figure*}

We start with our approved HST program (id:16301) and search for QSO sight lines in the Mikulski Archive for Space Telescopes (as of November 2022) within a projected radius of 1000 kpc from the Sculptor group, which is at (GLON,GLAT) $\approx (358.3\degrees, -77.3\degrees)$ and $d\sim3.9$ Mpc. We choose this region because it is an area of interest with numerous nearby galaxies, \HI\ clouds and the MS. When available, we use the coadded QSO spectra from the Hubble Spectroscopic Legacy Archive (HSLA; \citealt{Peeples17}). For sight lines that are not included or only partially included in HSLA, we download the CalCOS calibrated data files from the MAST archive and coadd the spectra using a published IDL package \texttt{coadd\_x1d.pro} \citep{danforth_coadd} over the wavelength range of the G130M and G160M gratings. The SNR of each spectrum is averaged across the full wavelength of the spectrum, same as the HSLA, and we only select QSO spectra with a minimum SNR = 8 per resolution element. 

We follow the apparent optical depth method (AOD) outlined in \cite{Savage91}, and convert the continuum-normalized absorption line profiles into apparent optical depths per unit velocity $\tau_a(v) = -{\rm ln}[F(v) / F_c(v)]$, where $F_c(v)$ is the continuum level and $F(v)$ is the observed flux as a function of velocity. The apparent column density per unit velocity, $N_a(v)$, relates to $\tau_a(v)$:
\begin{equation}
N_a(v) = \frac{m_e c}{\pi e^2} \times \frac{\tau_a(v)}{f\lambda} = 3.768 \times 10^{14} \frac{\tau_a(v)}{f\lambda({\rm \AA)}} \rm{cm^{-2} (km\cdot s^{-1})^{-1}},
\label{eq:aod_column}
\end{equation}
where $f$ is the transition oscillator strength, $\lambda$ is the rest wavelength in \AA. We calculate the total column densities by integrating the line profiles over velocity intervals specific to individual absorbers, $N_a = \int_{v_{\rm{min}}}^{v_{\rm{max}}} N_a(v) dv$, where $v_{\rm{min}}$ and $v_{\rm{max}}$ are the boundaries of the absorption. We estimate the equivalent widths of the line profiles to confirm absorption detected at $> 3 \sigma$ level.

\section{Comparison with Peek et al. (2008)} \label{sec:peek_compare}
Identifying absorbers associated with the Magellanic System often relies on simple approaches that take into account the absorbers' proximity to the Magellanic System in both position and velocity space. In this section, we compare our association algorithm with another association technique defined by \cite{Peek_Dparam} that also quantitatively estimates the 2D distance between an individual cloud and nearby gases structures. 

\begin{figure}[h!]
    \centering
    \includegraphics[width=\linewidth]{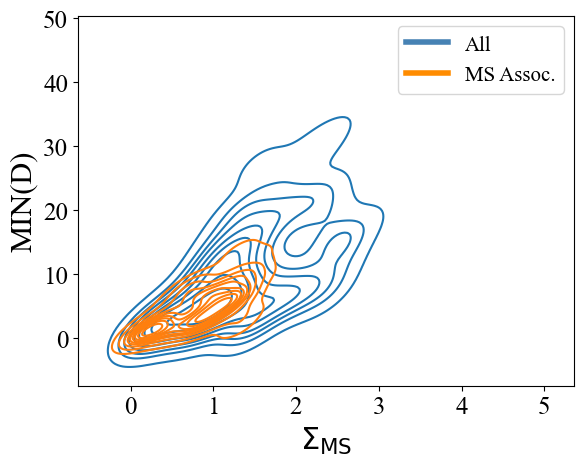}
    \caption{Comparison between the $\Sigma_{\rm{MS}}$ metric used in this work and the minimum values of the D parameters proposed by \cite{Peek_Dparam}. Blue contours represent all absorbers from Table \ref{tb:qso}, and the orange contours show MS-associated absorbers. Our $\Sigma_{\rm{MS}}$ metric is similar to the D parameters, but is specifically designed based on the spatial and kinematic distributions of the neutral and ionized gas in and surrounding the MS. We find a tighter correlation (orange contours) between the minimum D parameters and the $\Sigma_{\rm{MS}}$ values for MS-associated absorbers. 
    }
    \label{fig:Dcorr}
\end{figure}

To quantify associations between gaseous structures, \cite{Peek_Dparam} introduced a \textit{D parameter}, which is a distance metric in position-velocity space between clouds. The D parameter is defined as:
\begin{equation}
D = \sqrt{\delta \theta^2 + f^2 \cdot \delta v^2}, 
\label{eq:Dparam}
\end{equation}
where $\delta \theta$ and $\delta v$ are the angular distance and 
velocity offset between two clouds in Galactic Standard of Rest (GSR), and $f$ is a conversion factor that parameterizes the significance between angular and velocity distances. \cite{Peek_Dparam} chose $f$ to be 0.5$^{\circ}$/km$\cdot$ s$^{-1}$ based on the cloud clustering observed in simulated high-velocity cloud complexes, and concluded that clouds within the same complex typically exhibit $D<25^{\circ}$.

Figure \ref{fig:Dcorr} presents a comparison between \citeauthor{Peek_Dparam}'s D parameter and the $\Sigma_{\rm{MS}}$ values from our algorithm. We compute the minimum D parameters between the MS and all absorbers listed in Table \ref{tb:qso} and compare them with $\Sigma_{\rm{MS}}$. We also compare the mean and median D parameters among absorbers with $\Sigma_{\rm{MS}}$ and find similar results. Figure \ref{fig:Dcorr} shows that the MS absorbers yield a tighter correlation between the minimum D parameters and $\Sigma_{\rm{MS}}$ values (yellow contours), and they fall within the clustering limit of $D<25^{\circ}$ as established by \cite{Peek_Dparam}. A Pearson correlation coefficient ($r$ value) for all absorbers is 0.60, while the MS absorbers have a higher correlation coefficient ($r$=0.72) and a lower $p$ value. 

\startlongtable
\begin{deluxetable*}{cccccccccc}
\tabletypesize{\footnotesize}
\tablecaption{List of Selected Absorbers}
\tablehead{
\colhead{ID} \vspace{-0.2cm} & \colhead{QSO} & \colhead{GLON} & \colhead{GLAT} &  \colhead{Ion} & \colhead{Identified} & \colhead{v(min, max)} & \colhead{logN} &  \colhead{S/N} & \colhead{Reference} \\[0.05cm]
\colhead{} &  \colhead{} & \colhead{[deg.]} & \colhead{[deg.]} & \colhead{} & \colhead{Association} & \colhead{[km/s]} & \colhead{[log(cm$^{- 2}$)]} & \colhead{} & \colhead{}\\ 
\colhead{(1)} & \colhead{(2)} & \colhead{(3)} & \colhead{(4)} & \colhead{(5)}  & \colhead{(6)} & \colhead{(7)}& \colhead{(8)}& \colhead{(9)}& \colhead{(10)}} 

\startdata 
0 & 1H0419-577 & 266.987 & -41.996 & CII & MS & (100.0, 136.0) & $>$13.80 & 114.0 & R17\\
& & & & CIV	& MS & (104.0, 206.0) & 12.61$\pm$0.14 & 63.0 & R17 \\
& & & & SiII & MS & (100.0, 134.0) & 12.58$\pm$0.08 & 83.0 & R17 \\
& & & & SiII & MS & (159.7, 215.5) & 11.52$\pm$0.30 & & K22 \\
& & & & SiII & MS & (227.9, 287.9) & 12.63$\pm$0.03 & & K22 \\
& & & & SiIII & MS & (100.0, 129.0) & 12.18$\pm$0.05 & 69.0 & R17 \\
& & & & SiIII & MS & (159.7, 215.5) & 11.84$\pm$0.18 & & K22 \\
& & & & SiIII & MS & (227.9, 287.9) & 12.50$\pm$0.05 & & K22 \\
\enddata

\tablenotetext{}{Note: Col (1): Unique IDs assigned to individual QSO sight lines. Col (2): QSO Names. 
Cols (3) \& (4): QSO locations in Galactic coordinates. 
Col (5): Ions detected along QSO sight lines. 
Col (6): Primary associations of absorbers identified by our algorithm. ``MS" means the primary association is with the \HI\ in the MS, ``U(MS)" means secondary MS associations, ``U" means uncertain associations, and ``G" means associations with galaxies and the letters in the parentheses indicate the corresponding galaxies as listed in Table \ref{tb:gal_info}. We discuss the method to identify each absorber's association in Section \ref{sec:determine_assoc}.
Col (7): Minimum and maximum LSR velocity ranges for ion absorption.
Col (8): Ion column densities. \citetalias{lehner20_amiga} quotes a 2$\sigma$ upper limit on the column density in cases the line is not detected at 2$\sigma$ significance. 
Col (9): Signal$-$to-noise ratio (S/N) values (per resolution element), which are directly sourced from the original literature as listed in Col 10. 
\citetalias{richter17} calculated their S/N averaged over an absorption-free spectral range of 1208--1338 \AA, \citetalias{fox20} estimated theirs near \SII\ 1250 \AA, and \citetalias{lehner20_amiga} estimated theirs near \SiIII\ 1206 \AA. The S/N values of our archive sight lines are estimated over a wide spectral range of 1150--1750 \AA\ (see \citealt{zheng23}). 
As the \citetalias{Krishnarao22} sample did not provide S/N values for their published absorbers, 
we leave these entries empty.
Col (10): Data references, from which we adopt the ion measurements.} 
\tablecomments{Table 1 is published in its entirety in the machine-readable format.
      A portion is shown here for guidance regarding its form and content.}
\label{tb:qso}
\end{deluxetable*}

\newpage
\bibliographystyle{aasjournal}
\bibliography{main}

\end{document}